\pdfoutput=1
\PassOptionsToPackage{table}{xcolor}
\documentclass[sigconf]{acmart}
\usepackage[linesnumbered,vlined,algoruled]{algorithm2e}
\usepackage{setspace}
\usepackage{lipsum}
\usepackage[justification=centering]{caption}
\usepackage{subfigure}
\usepackage{wrapfig}
\usepackage{tikz}
\usepackage{tabularray}
\usepackage{multirow}
\usepackage{subcaption}

\AtBeginDocument{%
  }

\definecolor{skyblue}{RGB}{211,234,248} 

\newcommand{\R}[1]{{%
    \textbf{%
        \ifstrequal{#1}{1}{\textcolor{red}{R#1}}{%
        \ifstrequal{#1}{2}{\textcolor{blue}{R#1}}{%
        \ifstrequal{#1}{3}{\textcolor{magenta}{R#1}}{%
        \ifstrequal{#1}{4}{\textcolor{teal}{R#1}}{%
                           \textcolor{cyan}{R#1}%
        }}}}%
    }%
}}

\usepackage{hyperref}
\usepackage{listings}
\usepackage{xcolor}

\makeatletter
\def\UrlAlphabet{%
      \do\a\do\b\do\c\do\d\do\e\do\f\do\g\do\h\do\i\do\j%
      \do\k\do\l\do\m\do\n\do\o\do\p\do\q\do\r\do\s\do\t%
      \do\u\do\v\do\w\do\x\do\y\do\z\do\A\do\B\do\C\do\D%
      \do\E\do\F\do\G\do\H\do\I\do\J\do\K\do\L\do\M\do\N%
      \do\O\do\P\do\Q\do\R\do\S\do\T\do\U\do\V\do\W\do\X%
      \do\Y\do\Z}
\def\UrlDigits{\do\1\do\2\do\3\do\4\do\5\do\6\do\7\do\8\do\9\do\0}
\g@addto@macro{\UrlBreaks}{\UrlOrds}
\g@addto@macro{\UrlBreaks}{\UrlAlphabet}
\g@addto@macro{\UrlBreaks}{\UrlDigits}
\makeatother

\definecolor{codegreen}{rgb}{0,0.6,0}

\lstdefinestyle{bashstyle}{
    commentstyle=\color{codegreen},
    basicstyle=\ttfamily\footnotesize,
    breaklines=true,
    showstringspaces=false,
    columns=fullflexible,
    keepspaces=true,
    frame=none, 
    backgroundcolor=  
}

\definecolor{color1}{rgb}{0.1,0.7,0.8} 
\definecolor{color2}{rgb}{0.9,0.1,0.1} 
\definecolor{color3}{rgb}{0.7,0.3,0.7} 
\definecolor{color4}{rgb}{0.3,0.3,0.7} 
\definecolor{color5}{RGB}{8, 102, 3} 
\definecolor{color6}{rgb}{0.53, 0.66, 0.42} 

\newcommand{\name}{\textsf{EdgeLoRA}}
\newcommand{\llama}{\textsf{llama.cpp}}
\newcommand{\eg}{\textit{e.g.}, }
\newcommand{\ie}{\textit{i.e.}, }

\setcopyright{acmlicensed}
\copyrightyear{2025}
\acmYear{2025}
\setcopyright{cc}
\setcctype{by}
\acmConference[MobiSys '25]{The 23rd Annual International Conference on Mobile Systems, Applications and Services}{June 23--27, 2025}{Anaheim, CA, USA}
\acmBooktitle{The 23rd Annual International Conference on Mobile Systems, Applications and Services (MobiSys '25), June 23--27, 2025, Anaheim, CA, USA}
\acmDOI{10.1145/3711875.3729141}
\acmISBN{979-8-4007-1453-5/2025/06}

\begin{document}

\title{EdgeLoRA: An Efficient Multi-Tenant LLM Serving System on Edge Devices}









\author{Zheyu Shen, Yexiao He, Ziyao Wang, Yuning Zhang, Guoheng Sun, Wanghao Ye, Ang Li}
\affiliation{%
  \institution{University of Maryland, College Park}
  \city{College Park, MD}
  \country{USA}}
\email{{zyshen, yexiaohe, ziyaow, yuning, ghsun, wy891, angliece}@umd.edu}


\begin{abstract}
Large Language Models (LLMs) have gained significant attention due to their versatility across a wide array of applications. Fine-tuning LLMs with parameter-efficient adapters, such as Low-Rank Adaptation (LoRA), enables these models to efficiently adapt to downstream tasks without extensive retraining. 
Deploying fine-tuned LLMs on multi-tenant edge devices offers substantial benefits, such as reduced latency, enhanced privacy, and personalized responses. 
However, serving LLMs efficiently on resource-constrained edge devices presents critical challenges, including the complexity of adapter selection for different tasks, memory overhead from frequent adapter swapping. Moreover, given the multiple requests in the multi-tenant settings, processing requests sequentially will result in underutilization of computational resources and significant latency.
This paper introduces \name, an efficient system for serving LLMs on edge devices in multi-tenant environments. 
\name\ incorporates three key innovations: (1) an adaptive adapter selection mechanism to streamline the adapter configuration process; (2) heterogeneous memory management, leveraging intelligent adapter caching and pooling to mitigate memory operation overhead; and (3) batch LoRA inference, which enables efficient batch processing to significantly reduce computational latency.
Comprehensive evaluations using the Llama3.1-8B model demonstrates that \name\ significantly outperforms the status quo (\ie \textsf{llama.cpp}) in terms of both latency and throughput. The results demonstrates \name\ could achieve up to 4$\times$ boost in throughput with less energy consumption. Even more impressively, it manages to serve several orders of magnitude more adapters simultaneously without sacrificing inference performance.
These results highlight \name’s potential to transform edge deployment of LLMs in multi-tenant scenarios, offering a scalable and efficient solution for resource-constrained environments.
\end{abstract}



\maketitle

\section{Introduction}\label{sec:introduction}

Large language models (LLMs) have emerged as transformative tools in natural language processing (NLP), excelling in diverse applications such as chatbots~\cite{chatbot,chatbot-luo}, summarization~\cite{abstractive-summary, summarization-survey, summarization-review}, code generation~\cite{code-llama, chatgpt-code}, and captioning~\cite{caption-generation}.
Innovations from Anthropic~\cite{claude}, Google~\cite{gemini}, Meta~\cite{llama,llama3.1}, Microsoft~\cite{phi3}, and OpenAI~\cite{gpt3,gpt4} demonstrate their ability to reframe tasks like translation~\cite{machine-translation}, classification~\cite{text-classification}, and question-answering~\cite{benchmark-qa} as language generation problems, achieving remarkable improvements.
Beyond text, LLMs extend to image~\cite{vit, clip-gen}, video~\cite{llm-video}, speech~\cite{llasm}, and multimodal~\cite{mmllm, multimodal-survey} generation. Their success is driven by the Transformer architecture~\cite{transformer} and its variants~\cite{swin, palm}, which effectively model complex relationships in sequences, outperforming traditional methods~\cite{lstm}. These advancements position LLMs as foundational to modern AI, transforming research and applications across domains.

Although pretrained LLMs already demonstrate superior performance on general tasks, they can be further enhanced for domain-specific applications through fine-tuning.
This adaptation process refines a pretrained LLM for optimal performance across diverse, specialized tasks. 
This pretrain-then-finetune paradigm has led to the proliferation of numerous fine-tuned variants of a single base LLM, each tailored to a specific task (\eg dialogue~\cite{instructdial}, writing~\cite{writing-alpaca}, and code generation\cite{code-llama, chatgpt-code}) or domain (\eg medical~\cite{chatdoctor}, mathematics~\cite{mathscale, goat}, and legal~\cite{lawllm}).
To address the computational overhead of fine-tuning, parameter-efficient methods like Low-rank adaptation (LoRA) have been developed~\cite{lora, qlora}. 
LoRA reduces computational demands by updating only a small portion of model parameters, which exploits the low dimensionality of parameter updates in fine-tuning and representing them with pairs of low-rank matrices, \ie LoRA adapters. Compared to fully fine-tuning, LoRA can reduce the number of training parameters by $10,000\times$ while maintaining comparable performance~\cite{lora}. 

\begin{figure}[h]
  \centering
  \includegraphics[width=\linewidth]{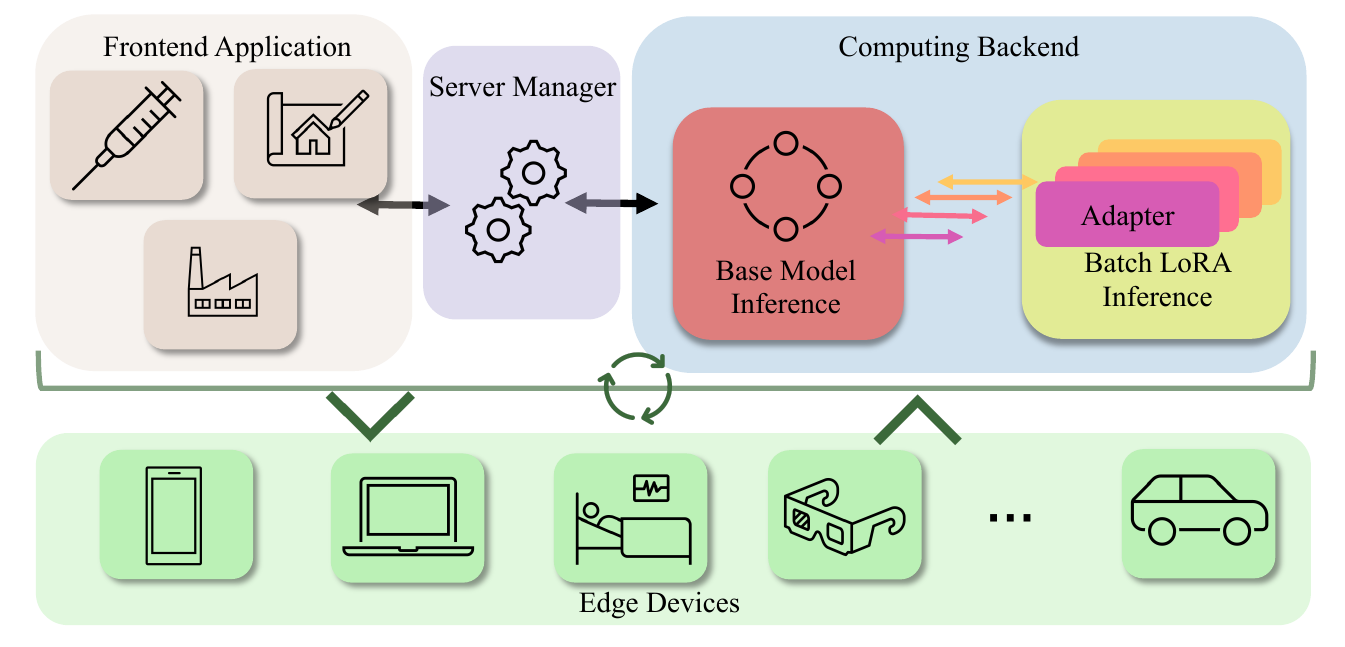}
  \caption{Multi-tenant LLM Serving on Edge Devices.}\label{fig:intro}
\end{figure}

The use of multi-tenant LLM applications on edge devices has grown significantly, spanning diverse domains such as personalized virtual assistants~\cite{personal-agent}, real-time translation tools~\cite{translation}, context-aware chatbots~\cite{chatbot, chatbot-luo}, and on-device content moderation~\cite{writing-alpaca,writing-assistant}. These applications highlight the demand for flexible, efficient, and scalable serving strategies tailored for edge environments.
To enable personalization and task-specific optimization, one promising solution involves serving a set of LoRA models fine-tuned on a shared base LLM on edge devices, as illustrated in Figure~\ref{fig:intro}.
As presented in LoRA~\cite{lora}, merging LoRA adapters into the pretrained weights eliminates additional latency overhead during inference.
However, serving multiple LoRA models concurrently introduces complexity, \ie adapters can be swapped by adding or subtracting LoRA matrices from the base model, but this approach incurs significant latency overhead and drastically reduces throughput. 
Alternatively, inference can proceed without merging by processing the base model and LoRA adapters in parallel, though this method also presents inefficiencies.
Despite these approaches, serving multiple LoRA adapters with a shared base model on edge devices remains challenging:
(1) users must manually select the appropriate adapter from a large pool, a process that is cumbersome and prone to errors; 
(2) efficient memory management is essential, particularly for swapping adapters between memory and disk to minimize resource overhead while increasing the number of adapters hosted by a single device; 
and (3) unmerged LoRA inference incurs substantial latency, as the parallel computation capabilities of edge devices are often underutilized when managing dynamic workloads.  These challenges are particularly pronounced in the context of multi-tenant edge devices, which must efficiently manage multiple fine-tuned models concurrently. 

\noindent
\textbf{Status Quo and Its Limitations.} 
Current LLM serving systems, such as vLLM~\cite{vllm} and SGLang~\cite{sglang}, are primarily designed to serve pretrained models efficiently. Some approaches, like SLoRA~\cite{slora}, mitigate memory constraints by dynamically swapping LoRA adapters between memory and disk, avoiding the need to load all adapters into memory simultaneously.
Similarly, dLoRA~\cite{dlora} reduces latency by scheduling merged and unmerged LoRA inference operations. 
Other solutions, such as Punica~\cite{punica}, improve GPU utilization by batching pretrained weights during LoRA inference to enable parallelism.
Despite these advancements, most existing solutions require users to manually select appropriate adapters, presenting a significant obstacle for users unfamiliar with the capability of adapters.  
Moreover, these systems fail to address the distinct challenges posed by multi-tenant edge devices. Unlike server environments, edge devices operate under severe resource constraints, highly dynamic workloads, and require support for heterogeneous system architectures, such as CPU, Metal, or BLAS-based backends. 
One serving framework tailored for edge devices is \llama~\cite{llamacpp}, which provides versatile support for multiple computing backends. However, \llama\ lacks efficient support for LoRA inference, as it can only process requests that use the same adapters simultaneously. This limitation restricts its applicability in multi-tenant scenarios, where diverse adapters must be managed concurrently.
These limitations highlight the pressing need for an efficient LLM serving system specifically designed to address the unique challenges of multi-tenant edge environments.

Efficiently serving LLMs on multi-tenant edge devices presents challenges such as selecting LoRA adapters, managing memory overhead, and optimizing efficiency. Specifically, adapter selection can be complex, as it requires accurately matching user requests to suitable adapters from potentially large and diverse adapter pools, making manual selection cumbersome and error-prone. Memory management becomes critical due to the limited resources available on edge devices; swapping adapters between memory and disk must be handled carefully to prevent throughput degradation and excessive latency. Additionally, optimizing computational efficiency is challenging, given that unmerged LoRA inference operations can lead to significant latency, particularly when parallel processing capabilities are underutilized, resulting in suboptimal throughput on dynamically changing workloads.

\noindent
\textbf{Overview of the Proposed Approach. } 
In this work, we introduce \name, an efficient multi-tenant LLM serving system designed specifically for edge devices to address the key challenges associated with serving multiple LoRA adapters.
To eliminate the need for users to manually specify adapters while maintaining performance, 
\name\ employs an adaptive adapter selection to automatically identify and deploy the optimal adapter based on request-specific requirements and the availability of adapters in the memory cache. 
To reduce the overhead associated with the swapping of adapters on resource-constrained edge devices, \name\ incorporates a heterogeneous memory manager that utilizes both the memory cache and a pre-allocated memory pool for efficient adapter management. 
Furthermore, \name\ introduces batch LoRA inference, a method that combines inference for pretrained weights and LoRA adapters into a unified process, significantly improving the utilization of computational resources under dynamic workloads.
By combining these components, \name\ ensures efficient resource utilization, reduced inference latency, and robust adaptability for multi-tenant edge environments.

\noindent
\textbf{System Implementation and Evaluation Results.}
We implemented \name\ with over 1k+ lines of C++ code, extending the \llama\ framework. For evaluation, \name\ was tested by serving Llama3.1-8B, Llama3.2-3B, and OpenELM-1.1B on three representative edge devices: Jetson AGX Orin, Jetson Orin Nano and Raspberry Pi 5. 
Experimental results demonstrate that \name\ significantly outperforms the state-of-the-art inference library for edge devices, \llama. Specifically, \name\ achieves throughput improvements of 2-4$\times$ across a variety of tasks while increasing the number of concurrently served adapters by several orders of magnitude, all without compromising inference performance.

\noindent
\textbf{Summary of Contributions.}
The key contributions of \name\ are summarized as follows:

\begin{itemize}
    \item \name\ introduces an adaptive adapter selection mechanism that dynamically selects the most appropriate adapter based on incoming request requirements and memory availability, reducing manual intervention and adapter switching overhead while ensuring robust task performance.
    \item To optimize memory usage on edge devices, \name\ employs a hybrid memory management strategy that combines caching and pre-allocated memory pools, minimizing allocation overhead and reducing latency.
    \item \name\ improves GPU utilization by batching requests with different adapters, enabling simultaneous computation for base model inference and adapter-specific weights, thereby enhancing throughput and efficiency.
\end{itemize}
\section{Background and Motivation}\label{sec:background}


\subsection{Low-Rank Adptation}\label{subsec:peft}

\begin{figure}[h]
  \centering
  \subfigure[LoRA fine-tuning.]{\includegraphics[width=0.55\linewidth]{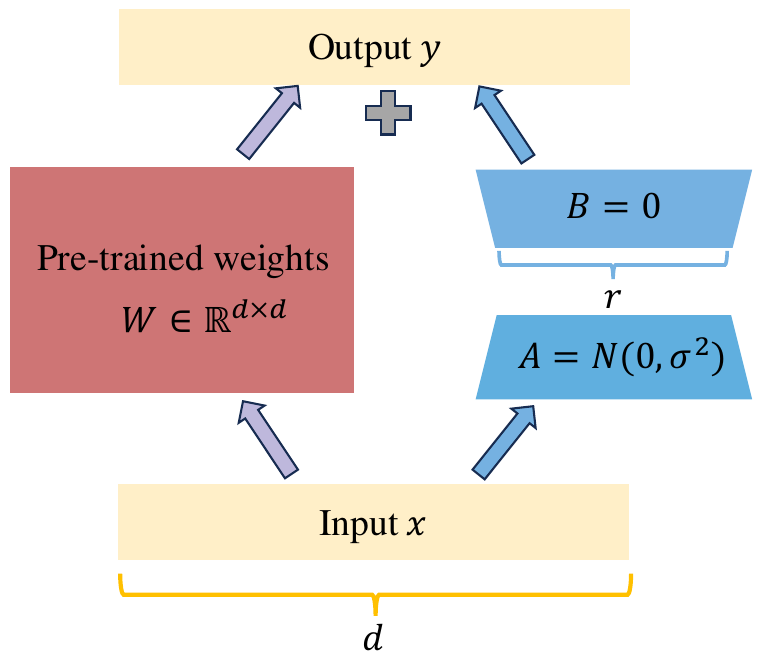}\label{subfig:lora}}
  \subfigure[LoRA inference.]{\includegraphics[width=0.38\linewidth]{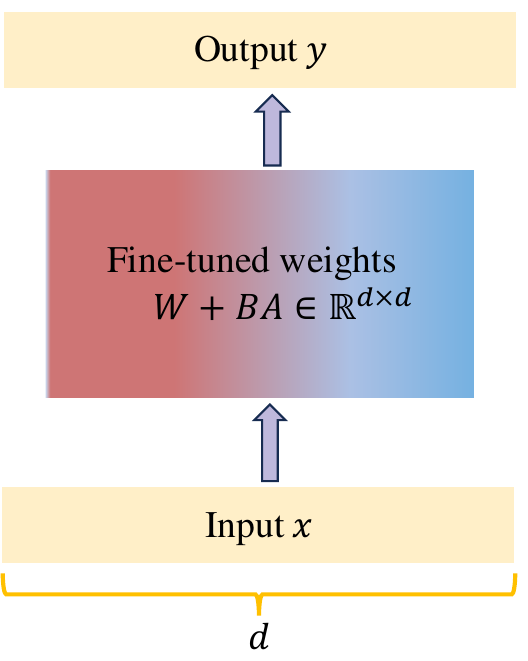}\label{subfig:lora_inference}}
  \caption{The workflow of LoRA.}\label{fig:lora}
\end{figure}

LoRA~\cite{lora, qlora, glora} is a parameter-efficient fine-tuning (PEFT)~\cite{peft-library} technique that enables the adaptation of LLMs to new tasks without requiring full model re-training. The motivation for developing LoRA arises from the observation that weight updates during adaptation exhibit a low intrinsic dimensionality. Specifically, as Figure~\ref{subfig:lora} illustrates, LoRA retains the pre-trained base model weights \( W \in \mathbb{R}^{d \times d} \) in a frozen state and augments each layer with trainable low-rank matrices \( A \in \mathbb{R}^{r \times d} \) and \( B \in \mathbb{R}^{d \times r} \) during the training phase, where \( r \ll d \). This approach enables a significant reduction in the number of trainable parameters, thereby substantially decreasing memory consumption.

Compared to traditional full-parameter fine-tuning approaches, LoRA achieves a reduction in trainable parameters by several orders of magnitude (up to a 10,000$\times$) while maintaining comparable performance in terms of model accuracy. During the inference phase, as illustrated in Figure~\ref{subfig:lora_inference}, LoRA merges the product of the matrices \( B \times A \) with the original weight matrix \( W \), effectively incorporating the weight updates \( \Delta W \) without incurring additional computational overhead. This characteristic distinguishes LoRA from earlier adapter-based methods~\cite{adapter-peft} and prompt-tuning~\cite{prompt-peft}, which often introduce extra latency during inference.
Due to its significant reduction in training and weight storage costs, LoRA has been widely adopted by the research community. Furthermore, LoRA has been employed extensively to enhance the capabilities of LLMs, such as in applications involving long-sequence modeling~\cite{longlora} and multimodal input~\cite{multimodalgpt}.

\subsection{Serving LLM on Edge Devices}

Most LLMs are built upon the Transformer architecture~\cite{transformer}, with the number of parameters ranging from several billion to several trillion~\cite{gpt3, llama, mixtral}.
These models aim to predict the next token conditioned on all preceding tokens, operating via an autoregressive generation process. 
During inference, tokens are iteratively generated based on the initial prompt and previously generated tokens until an end-of-sequence marker is reached. 
This autoregressive nature, coupled with the large parameter size, results in two key challenges for LLM inference: 1) variable inference latency, which depends on both input and output sequence lengths; and (2) high memory consumption, due to the need to maintain intermediate states for each active request.

The majority of popular LLM serving frameworks, such as vLLM~\cite{vllm}, Text Generation Inference~\cite{textgeninference}, and DeepSpeed-MII~\cite{deepspeedmii}, are designed primarily for x86 operating systems and CUDA backends. However, edge devices are characterized by diverse operating systems and computational backends, such as CPU, Metal, and BLAS. The \textsf{llama.cpp} framework~\cite{llamacpp} provides a versatile LLM serving solution, predominantly written in C++, and supports multiple backends through the use of the GGML tensor library~\cite{ggml}. \textsf{llama.cpp} addresses the challenge of dynamic incoming requests using a slot state machine, which groups all available tokens across requests into a single batch, thereby enabling parallel processing of requests. Additionally, \textsf{llama.cpp} mitigates memory consumption by employing model quantization techniques, making it one of the most popular LLM serving frameworks for edge devices.


\subsection{Fine-Tuned Adapters for Specialized Applications}

Pre-trained LLMs~\cite{gpt3, llama3.1, mixtral} have demonstrated remarkable capabilities in solving a variety of general tasks by leveraging their extensive pretraining on large and diverse datasets. 
These models are proficient at capturing the nuances of language, encoding semantic relationships, and handling a broad spectrum of use cases. However, despite their general-purpose strengths, pre-trained models often fall short when applied to specialized domains requiring unique language, terminology, or domain-specific knowledge.

To bridge this gap, fine-tuning pre-trained models on targeted datasets tailored to specific applications has become a widely adopted approach.
For example, Writing-Alpaca-7B~\cite{writing-alpaca}, fine-tuned from LLaMA-7B~\cite{llama} using a writing instruction dataset (an extension of the EDITEVAL~\cite{editeval-dataset} benchmark), significantly enhances LLaMA’s performance in writing tasks. Writing-Alpaca-7B consistently outperforms larger off-the-shelf LLMs in tasks requiring advanced writing assistance.
Similarly, ChatDoctor~\cite{chatdoctor} is based on the fine-tuned LLaMA-7B~\cite{llama} model, utilizing the Alpaca instruction dataset~\cite{alpaca} combined with the HealthCareMagic100k patient-doctor dialogue dataset. This fine-tuning enables ChatDoctor to better understand patient concerns and deliver informed, domain-specific advice, surpassing the performance of generic LLMs in medical consultation tasks.

\begin{table}[]
\caption{\label{tab:adapter_forget}
    OpenMath2-8B outperforms Llama3.1-8B-Instruct on math related tasks. But it sacrifices on other general-purpose tasks. 
}
\begin{tabular}{l|cc}
\toprule
\textbf{Task}     & \textbf{Llama3.1-8B-Instruct} & \textbf{OpenMath2-8B} \\ \hline
GSM8K    & 84.5        & 91.7         \\
MATH     & 51.9        & 67.8         \\ \hline
MMLU   &  68.2          &  36.9            \\
MMLU-PRO      &  37.9           &  16.3            \\
IFEVAL & 41.8            & 17.2             \\
\bottomrule
\end{tabular}
\end{table}

However, a key challenge associated with the fine-tuning process is the trade-off between generalization~\cite{lora-forget} and specialization. Fine-tuning a model for a specific domain often leads to a decline in its performance on tasks outside that domain. 
For example,  OpenMath2-8B~\cite{openmath2}, fine-tuned on Llama3.1-8B-Base~\cite{llama3.1} using the OpenMathInstruct-2~\cite{openmath2} dataset, achieves exceptional performance in mathematical tasks. In contrast, Llama3.1-8B-Instruct~\cite{llama3.1} also fine-tuned on Llama3.1-8B-Base using the general-purpose datasets, maintaining a broader range of capabilities. 
As shown in Table~\ref{tab:adapter_forget}, OpenMath2-8B surpasses Llama3.1-8B-Instruct on the whole MATH benchmark~\cite{math-dataset} by $15.9\%$. However, this improvement comes at the expense of reduced general-purpose capabilities compared to the pretrained model. This highlights the challenge of identifying a universally optimal adapter for serving LoRA models. Instead, adapter selection must be carefully tailored to align with the specific requirements of the application, ensuring that the chosen adapter delivers the desired balance between specialization and generalization.

\begin{figure*}
  \centering
  \includegraphics[width=.85\linewidth]{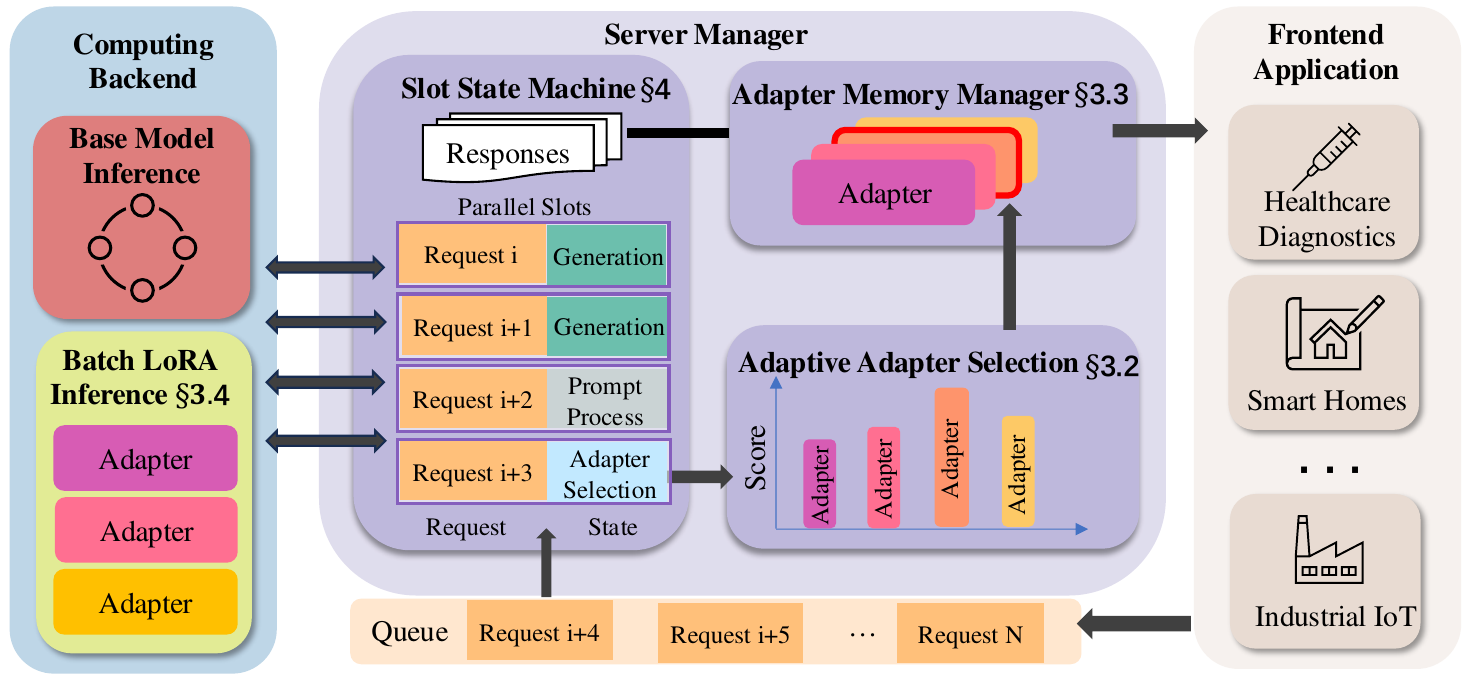}
  \caption{Overview of \name\ design.}\label{fig:overview}
\end{figure*}

\section{System Design}\label{sec:system_design}


\subsection{Overview}
\name\ introduces an efficient multi-tenant serving system specifically designed for deploying LLMs with multiple LoRA adapters on resource-constrained edge devices. 
The system addresses three critical challenges: dynamically selecting suitable adapters, efficiently managing adapter memory, and improving the efficiency of LoRA inference for concurrent requests.
Figure~\ref{fig:overview} provides a high-level view of the \name. 
As our implementation is based on \llama, \name\ shares a similar design structure, comprising two main components: the \textbf{Server Manager} and the \textbf{Computing Backend}.
The Server Manager oversees request handling and memory management, with its core functionality driven by the \textbf{Slot State Machine} (\S \ref{sec:implementation}), which manages concurrent requests. This component is crucial for dynamically allocating system resources, selecting appropriate adapters, and ensuring efficient operations under multi-tenant workloads. Specifically, the Server Manager includes two key modules: \textbf{Adaptive Adapter Selection} (\S \ref{subsec:adapter_selection}), responsible for intelligent adapter selection, and the \textbf{Adapter Memory Manager} (\S \ref{subsec:memory_manage}), which ensures automatic adapter selection and  efficient memory usage. Once requests are grouped into batches and the required LoRA adapters are loaded into the memory cache, the \textbf{Computing Backend} constructs and executes the computational graph. The Computing backend integrates the base model inference with \textbf{Batched LoRA Inference} (\S \ref{subsec:batch_lora}) to optimize the resource utilization and hence improve efficiency.

\begin{figure}[h]
  \centering
  \includegraphics[width=\linewidth]{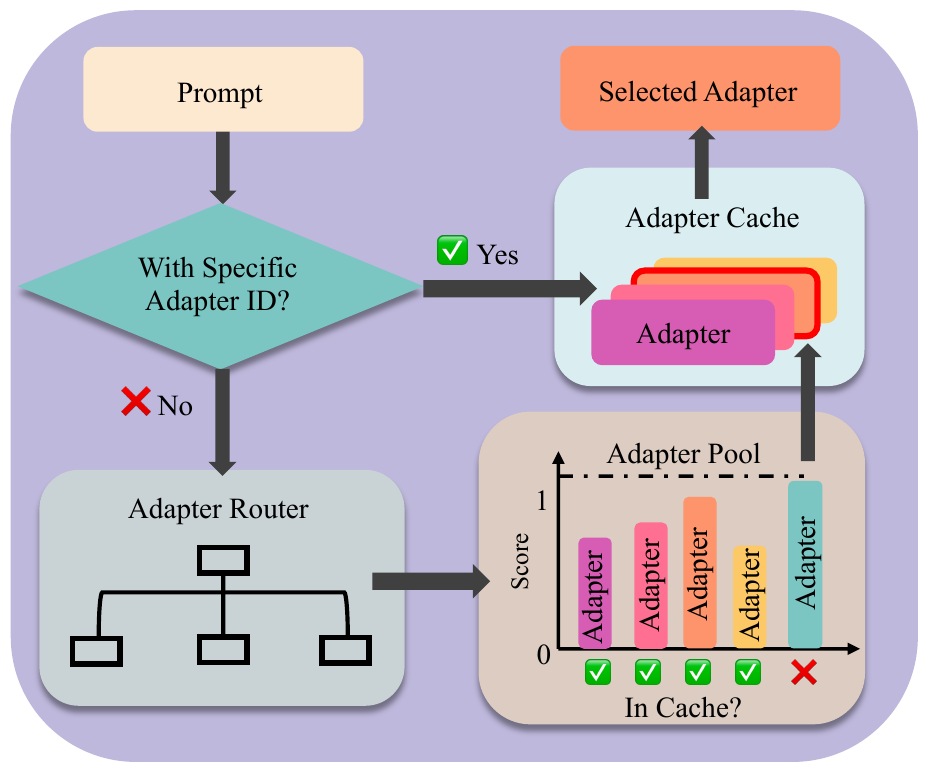}
  \caption{The workflow of adaptive adapter selection.}\label{fig:adaptive_adapter_selection}
\end{figure}

\subsection{Adaptive Adapter Selection}\label{subsec:adapter_selection}
Selecting the optimal LoRA adapter for a given request poses significant challenges due to the variability in application requirements and the computational overhead of loading and switching adapters. To address these challenges, \name\ introduces an adaptive adapter selection mechanism. This mechanism dynamically identifies and deploys the optimal adapter based on application-specific needs and the availability of adapters in the memory cache.
As illustrated in Figure~\ref{fig:adaptive_adapter_selection}, the adaptive selection mechanism operates as follows: (1) when a new request is received, the system first checks if a specific adapter ID has been provided. If an adapter is explicitly specified, it is directly employed, bypassing the adaptive selection process. (2) Otherwise, the adaptive selection mechanism is employed to choose the most suitable adapter by analyzing the incoming prompt and the availability of adapters in the memory cache.
The detailed steps of this mechanism are outlined in Algorithm~\ref{alg:adaptive_adapter_selection}.

\begin{algorithm}
\caption{Adaptive Adapter Selection}
\label{alg:adaptive_adapter_selection}
\KwIn{User prompt $x$, Memory cache $M$, Adapter set $A$, Evaluation datasets $D = \{d_1, d_2, \dots, d_m\}$, Adapter router $C$}
\KwOut{Selected Adapter $a^*$}

\If{Adapter $a$ is explicitly specified for request $x$}{
    \textbf{return} $a$  \tcp*{Bypass adaptive selection}
}

\If{Adapter router $C$ is not available}{
    \ForEach{dataset $d_i \in D$}{
        \ForEach{adapter $j \in A$}{
            Evaluate performance $P_{ij}$ for adapter $j$ using dataset $d_i$;
        }
    }
    
    Train adapter router $C$ with prompt $x$ as input and performance $P_{ij}$ for adapter $j$ using dataset $d_i$ as output; 
}

Compute confidence scores $S = \{s_1, s_2, \dots, s_n\}$ using adapter router $C$ for prompt $x$;

$A' \leftarrow$ Top-$k$ adapters from $A$ based on scores in $S$;

\ForEach{adapter $a' \in A'$ in descending order of confidence}{
    \If{$a' \in M$}{
        \textbf{return} $a'$  \tcp*{Adapter is available}
    }
}

Load the adapter with the highest score from $A'$ into memory cache $M$;

\textbf{return} adapter $a^*$ from $A'$;

\end{algorithm}

\textbf{Profiling-Based Adapter Selection.}
\name\ employs a prof-iling-based method to generate training data for an adapter router, wherein the performance of each adapter is evaluated using diverse public evaluation datasets. Let $D = \{d_1, d_2, \dots, d_m\}$ denote the set of evaluation datasets, and for each dataset $d_i$, the system determines the performance $P_{ij}$ for each adapter $j$. The best-performing adapters for each dataset are identified, and this information is used to train a multi-label classifier that serves as the adapter router.
In the adapter router, the user-provided prompt $x$ serves as the input, and the output comprises scores $s_j \in [0, 1]$ for each adapter $j$, indicating its suitability for the given request. After training on such profiling data, the adapter router can assign a score to each adapter based on the given prompt, thereby enabling informed decisions regarding adapter selection. Specifically, the adapter with the highest score, $\text{argmax}_j (s_j)$, will be chosen as the most suitable when considering the quality of the response context. 
To further optimize resource utilization, the system selects a subset of top-scoring adapters, $A' \subset A$, where $A$ is the full set of adapters.
The selected adapters are then checked for availability in the memory cache in descending order of confidence score. If any of these adapters are already in the cache, they are immediately employed for inference. If none of the selected adapters is cached, the adapter with the highest score is dynamically loaded for use.

This adaptive approach addresses the challenge of manually identifying the best adapters for LLM requests, especially in multi-tenant edge environments. By automating the selection process, it significantly reduces the overhead associated with switching adapters between memory and disk. This approach maximizes memory utilization while ensuring that high-confidence adapters are prioritized.

\subsection{Heterogeneous Memory Managemer}\label{subsec:memory_manage}

\begin{figure}[h]
  \centering
  \includegraphics[width=\linewidth]{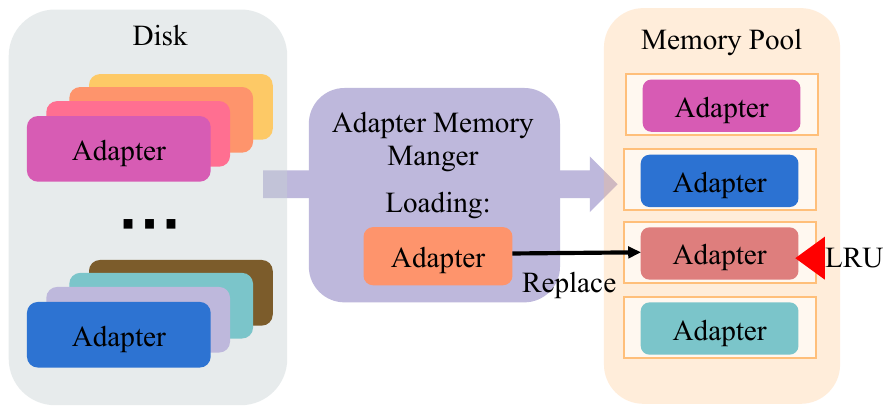}
  \caption{The adapter memory manager evicts the least frequently used adapter and loads the newly required one into a free memory block in the pool.}\label{fig:memory_manager}
\end{figure}

Edge devices often operate under strict memory constraints, posing significant challenges for serving LLM with multiple adapters concurrently. To address this limitation, \name\ incorporates a heterogeneous memory manager that utilizes both a memory cache and pre-allocated cache pools to effectively manage memory utilization. Figure~\ref{fig:memory_manager} illustrates the comprehensive architecture of Heterogeneous Memory Management.
Traditional LLM inference systems load all adapters into memory simultaneously and activate them as needed during computation. However, our approach optimizes memory usage by loading adapters only when required, thus significantly reducing the memory overhead. 

To minimize the frequent swapping of adapters between memory and disk, the system employs a memory cache $C$ to store frequently accessed adapters. Let $C = \{a_{c_1}, a_{c_2}, \dots, a_{c_l}\}$ be the set of cached adapters, where $l \leq k$. When the cache reaches its capacity, a replacement policy is used to evict the least frequently used and load the newly required adapter. This replacement policy aims to maximize the cache hit rate,  defined as $H = \frac{h_{\text{cache}}}{h_{\text{total}}}$, where $h_{\text{cache}}$ is the number of adapter requests successfully served from the cache, and $h_{\text{total}}$ is the total number of adapter requests.

To further reduce runtime memory allocation overhead, the heterogeneous memory manager employs a pre-allocated memory pool $P = \{p_1, \dots, p_l\}$, consisting of fixed memory blocks. 
Each memory block $p_i \in P$ corresponds to the size of a single adapter in the memory cache. When a new adapter is required, it is assigned to a free block $p_i$, avoiding dynamic memory allocation during runtime. This approach minimizes memory fragmentation, reduces the latency associated with memory allocation and deallocation, and enhances system stability under dynamic workloads.
By combining memory caching and pre-allocated memory pools, the heterogeneous memory manager ensures efficient memory utilization while maintaining low latency and high system stability. When used in cooperation with the adaptive adapter selection mechanism, this memory management approach significantly reduces memory overhead and ensures efficient operation in resource-constrained edge environments. 

\subsection{Batch LoRA Inference}\label{subsec:batch_lora}

Different from merged LoRA inference as illustrated in Figure~\ref{subfig:lora_inference}, unmerged LoRA inference separates the computations of the pre-trained LLM weights and LoRA adapter weights, allowing for more modular and dynamic computation during inference. 
Requests that require the same adapter can be grouped together, enabling shared computation for both the pre-trained LLM and LoRA adapter weights.
However, in practical multi-tenant environments, requests requiring the same adapter are relatively rare and dynamic workloads require simultaneous use of multiple LoRA adapters for diverse tasks, limiting opportunities to group them effectively during unmerged LoRA inference. To address this limitation, we propose a novel approach named Batch LoRA Inference, which enables requests with different adapters to be processed within a single batch, thereby maximizing computational efficiency.


\begin{figure}[h]
  \centering
  \includegraphics[width=\linewidth]{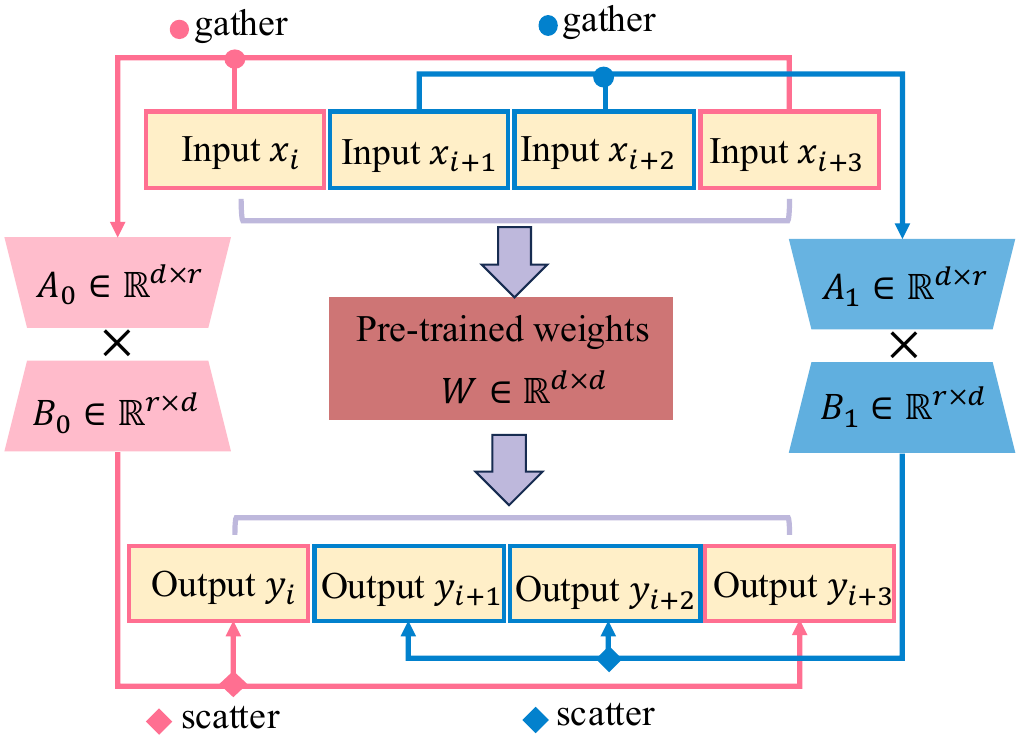}
  \caption{Batch LoRA inference. }\label{fig:batch_lora}
\end{figure}

Figure~\ref{fig:batch_lora} further illustrates batching LoRA inference. Consider a scenario with multiple distinct requests $\{x_0, \dots, x_n\}$, each requiring a unique LoRA adapter. 
During inference, the inputs for all requests are batched together, and the computations involving the pre-trained weights are performed in parallel,  represented as $W 	\times [x_0, x_1, \dots, x_n]$. 
To further optimize efficiency, requests requiring the same adapter are grouped into a unified batch (u-batch) for the computation of the LoRA components. This ensures that shared computations for the same adapter are processed efficiently, leveraging the parallelism of modern GPUs.
Once the computations involving the LoRA part are complete, the results from both the pre-trained part and the LoRA adapters are combined. Finally, the aggregated outputs are scattered back to their original locations in the input batch, yielding the final results, $[y_0, y_1, \dots, y_n] = W \times [x_0, x_1, \dots, x_n] + [B_0A_0x_0, B_1A_1x_1, \dots, B_nA_nx_n]$.

By batching requests together, the system is able to exploit the inherent parallelism of modern GPU architectures, thus reducing per-request latency and improving throughput. This improvement is especially beneficial in scenarios involving limited computing resources, where maintaining large batch sizes is critical for maximizing the utilization of computing resources. 
In addition, the ability to handle multiple adapters in a single batch allows for greater flexibility and adaptability, ensuring that computational resources are utilized efficiently in multi-tenant environments with diverse and dynamic workloads.
\section{Implementation}\label{sec:implementation}

\begin{figure}
  \centering
  \includegraphics[width=\linewidth]{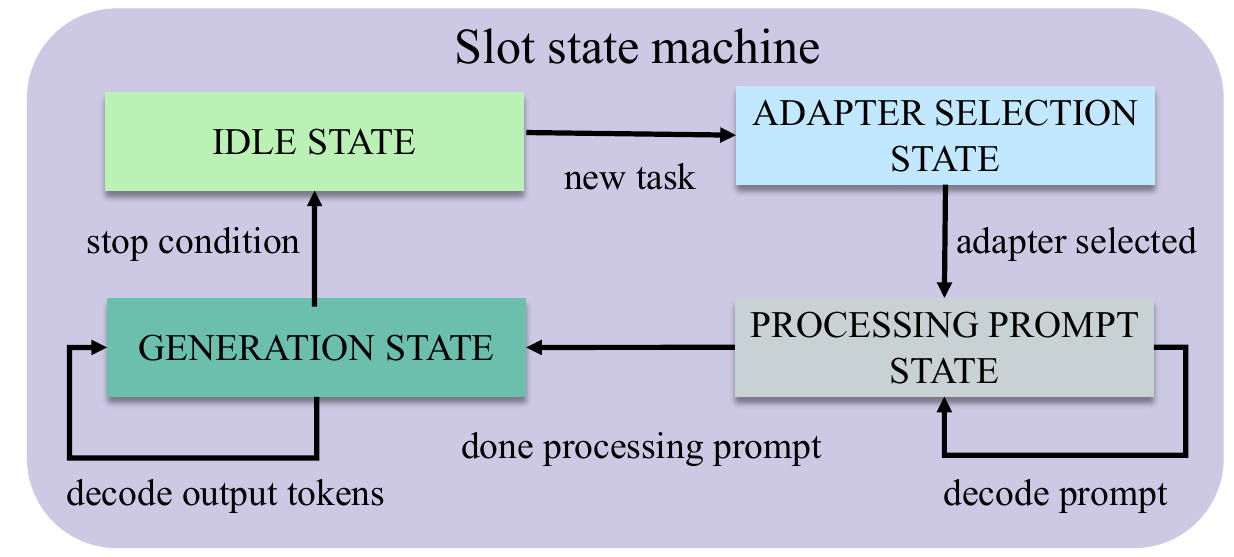}
  \caption{Slot state machine in our Server Manager.}\label{fig:state_machine}
\end{figure}

This section represents the implementation of \name\ based on \llama. \name\ consists of two key components: the Server Manager, which acts as the server front-end, and the Computing Backend, which handles inference operations. Together, these components address challenges related to adapter selection, memory management, and efficient LoRA inference.
The  Server Manager uses a slot state machine to manage multiple requests concurrently, as illustrated in Figure~\ref{fig:state_machine}. When a new request arrives, it is allocated to an idle slot, which transitions through the following states: (1) \textbf{Adapter Selection}, where Algorithm~\ref{alg:adaptive_adapter_selection} determines the optimal adapter for the request; (2) \textbf{Prompt Processing}, where the designated adapter is used to decode the input prompt; and (3) \textbf{Generation}, where tokens are decoded iteratively to generate the response. This design ensures efficient request handling while maintaining optimal resource utilization.
The Computing Backend processes the requests in batches forwarded by the Server Manager. It constructs a computational graph that integrates both base model inference and Batch LoRA Inference, optimizing computation efficiency for dynamic multi-adapter workloads.

The implementation of \name\ overcomes three primary challenges, including: (1) implementing a memory-efficient adapter router for precise task identification; (2) minimizing adapter-swapping overhead by implementing a robust cache management strategy; and (3) leveraging GPU parallelism during LoRA inference to maximize throughput.  
The system is implemented in over 1k+ lines of C++ code, with 500 lines of JavaScript simulating the front-end application.

\subsection{Adaptive Adapter Selection}

\name\ employs a memory-efficient adapter router fine-tuned with LoRA on the same base model deployed on edge devices. The router is implemented as a custom multi-label classifier using the HuggingFace Transformers Library~\cite{transformers-library}. A \texttt{Linear} layer is appended to the \texttt{LlamaModel}, with input dimensions corresponding to the model’s \texttt{hidden\_dim} and output dimensions representing the number of adapters. The loss function, \texttt{torch.nn.BCEWithLogits Loss}~\cite{pytorch}, is used to train the router, 
with ground-truth labels indicating which adapters can generate correct responses for given prompts.

To train the router, evaluations are conducted for each LoRA adapter on five benchmarks using the Eleuther AI Language Model Evaluation Harness~\cite{lm-eval}, including IFEval~\cite{ifeval}, BBH~\cite{bbh}, MATH~\cite{math-dataset}, GPQA~\cite{gpqa}, and MMLU-PRO~\cite{mmlu-pro}. The evaluation results are used as ground truth for training, with prompts processed consistently with the evaluation harness. 

In the adapter selection stage, the input prompt is processed by the adapter router, which consists of the shared base model and an additional \texttt{Linear} layer. Since the computational cost of the base model far exceeds that of the \texttt{Linear} layer, the overhead introduced by adaptive adapter selection is roughly equivalent to the time required for decoding the input prompt. This results in minimal additional computational overhead while enabling accurate adapter selection. Additionally, the adapter router efficiently leverages memory already occupied by the base model, incurring only negligible extra memory usage from the additional \texttt{Linear} layer.

\subsection{Heterogeneous Memory Management}

To minimize memory operation overhead while providing efficient and low-latency access to frequently utilized adapters, \name\ incorporates a heterogeneous memory management strategy combining a Least Recently Used (LRU)~\cite{lru} memory cache and a pre-allocated memory pool.

The memory cache is implemented utilizing an LRU policy to manage adapters effectively. The LRU cache retains frequently accessed adapters in memory, evicting less-used adapters when the cache is full. This aligns with the unbalanced locality of adapters observed in real-world scenarios~\cite{alpaserve}. The adapter invocation probabilities exhibit a long-tail distribution, with approximately $10\%$ of adapters accounting for roughly $80\%$ of the invocation probability~\cite{cara-serve}. When adapter locality becomes more unbalanced, some adapters are used more frequently. As a result, the LFU cache could achieve a higher cache hit rate, further improving the overall system throughput. The implementation employs the C++ Standard Template Library (STL)~\cite{stl}, specifically leveraging \texttt{std::list} and \texttt{std::unordered\_set} to implement the LRU policy. 
In the event that a new adapter must be loaded when the cache has reached its capacity, the least recently used adapter is evicted, and its resources are returned to the memory pool for future reuse. During server initialization, the memory cache is prefilled with random adapters. 

To reduce runtime overhead and mitigate the latency associated with frequent dynamic memory allocation, a pre-allocated memory pool has been implemented. This memory pool is comprised of memory blocks that are reserved during system initialization. The memory pool is represented by \texttt{std::stack<std::shared\_ptr<adapt er>>}, which keeps track of available memory blocks and enables efficient allocation and deallocation during runtime. By reusing these blocks, the system minimizes allocation latency and ensures stable memory management under dynamic workloads.
The combined utilization of an LRU cache and a pre-allocated memory pool enables efficient memory usage and provides low-latency access in resource-constrained environments. 

\subsection{Batch LoRA Inference}

This section presents the details of batching LoRA inference implementation. The baseline approach processes each sample independently for adapter-specific computations, while base model computations are batched across the entire input. Although batching base model weights improves computational efficiency, adapter-specific computations remain isolated for each sample, leading to suboptimal GPU utilization and higher latency, especially in diverse multi-adapter workloads.


To address these limitations, our proposed group LoRA computing batches computations for samples sharing the same adapter. The implementation maps adapter IDs to sample indices, gathers data into sub-batches, and performs LoRA-specific matrix multiplications in a single operation using optimized GPU routines. And the results will be scattered back to their original positions in the output tensor. Gather and scatter operations ensure data alignment, while maintaining efficiency for adapter-specific computations.

By batching both base model and adapter-specific computations, group LoRA computing fully exploits GPU parallelism, reducing redundant operations and minimizing per-sample processing overhead. This results in enhanced GPU utilization, lower latency, and higher throughput, particularly in prompt processing stage where multiple samples share the same adapter. This scalable approach is highly effective for real-world multi-adapter scenarios that demand efficient, low-latency processing.

\section{Evaluation}\label{sec:evaluations}
We evaluate the performance of \name\ on synthetic workloads. Specifically, we evaluate the scalability of \name\ by serving up to two thousand LoRA adapters simultaneously and compare it with \llama. We then perform ablation studies to verify the effectiveness of individual components.
\noindent
\textbf{Model.} 
We evaluate \name\ using three different models: Llama3.1-8B~\cite{llama3.1}, Llama3.2-3B, and OpenELM-1.1B~\cite{openelm}, which are popular open-source LLMs. Each model is paired with adapter, and quantization configurations are presented in Table~\ref{tab:exp_setting}. All LoRA adapters are quantized using the Q8\_0 format~\cite{ggml}.
 While we use these models for evaluation, \name\ is flexible and compatible with other transformer-based architectures, such as GPT-3~\cite{gpt3}, Phi3~\cite{phi3}, Mixtral MOE~\cite{mixtral}, and Qwen~\cite{qwen}.

\begin{table}[h]
\caption{\label{tab:exp_setting}
    Model, adapter, and quantization configurations.
}
\begin{tabular}{c|cccc}
\toprule
Setting & Base model  & LoRA rank & Quantization \\ \hline
S1      & Llama3.1-8B &  32           & Q8\_0           \\
S2      & Llama3.2-3B            &    16            &   Q4\_0           \\
S3        & OpenELM-1.1B            &    16           &    Q4\_0          \\
\bottomrule
\end{tabular}
\end{table}

\noindent
\textbf{Hardware.} 
We conduct our experiments across various edge devices, including Jetson Orin Nano (mid-tier), Jetson Orin AGX (high-tier), and Raspberry Pi 5. The Jetson devices are equipped with GPUs, while the Raspberry Pi relies solely on its CPU. These devices have memory capacities ranging from 8GB to 64GB. Our results demonstrate that \name\ can efficiently serve thousands LoRA adapters on resource-constrained edge devices.

\noindent
\textbf{Baselines.}
Since \llama\ is a comprehensive LLM serving system designed to support various edge devices and capable of simultaneously serving multiple LoRA models, and other frameworks either exclusively support server environments or do not support LoRA models like MLC-LLM~\cite{mlc-llm}, we compare several variants of \name\ against \llama~\cite{llamacpp}.
\begin{itemize}
    \item \llama\ is an LLM serving system implemented entirely in C++. It supports multiple computation backends, including CPU, GPU, and METAL. When serving multiple LoRA models, the system loads all LoRA models into memory during server initialization. Users can send requests to dynamically adjust the scaling of different LoRA models and deploy them for computation as needed.
    \item \name\ builds upon the full feature set of \name, incorporating dynamic adapter selection to automatically choose the most suitable adapter for each request.
    \item \name (w/o AAS) is a variant of \name\ where adaptive adapter selection is disabled, requiring all requests to manually specify an adapter.
\end{itemize}

\noindent
\textbf{Metrics.}
The performance of serving systems can be evaluated using several key metrics, including latency and throughput. Following common practice, we report \textit{throughput}, \textit{average request latency}, \textit{average first-token latency}, and \textit{SLO attainment}. SLO (Service Level Objective) attainment is defined as the percentage of requests that return the first token within 6 seconds. Additionally, we also evaluate \textit{power consumption}, which provides a quantitative analysis of the energy usage of edge devices.

\subsection{End-to-End Results on Synthetic Workloads}

\textbf{Workload trace.} 

We generate synthetic workload traces using a Gamma process to model the arrival intervals of requests. The total request rate across all adapters is $R$ requests per second. For $n$ adapters, the optimal adapter for requests are sampled according to a power-law distribution with exponent $\alpha$, determining adapter locality. Specifically, the probability $P(i)$ of selecting adapter $i$ with adapters sorted by frequency, is defined by $P(i) = \frac{i^{-\alpha}}{\sum_{j=1}^{n} j^{-\alpha}}$. This choice of a power-law distribution is motivated by the observed long-tail distribution of adapter invocation probabilities in real-world workloads~\cite{cara-serve}. A lower $\alpha$ leads to higher locality, meaning requests are concentrated on fewer adapters, while a higher $\alpha$ results in a more uniform distribution across adapters. Arrival intervals between consecutive requests follow a Gamma distribution characterized by a shape parameter ($1 / cv^2$) and a scale parameter ($cv^2 / R$), where the coefficient of variation ($cv$) controls workload skewness or burstiness. A higher $cv$ introduces greater variability and burstiness in request arrival patterns.
This approach accurately simulates dynamic workloads for benchmarking.
To simulate real requests and process adapter selection, after \name\ invokes the adapter router, we generate $ k $ ordered adapters, denoted as $ A' $. Also for \name and \name (w/o AAS), we set the number of slot as $\gamma$. Our tests evaluate various combinations of parameters, including $\gamma$, $k$, $\alpha$, $R$, and $cv$. For each request, the input and output lengths are sampled from uniform distributions $ U[I_l, I_u] $ and $ U[O_l, O_u] $, respectively. By default, each trace lasts for 5 minutes. To conduct comprehensive experiments, we first pick a set of default parameters for generating workloads, as shown in Table~\ref{tab:default_parameters}. 

\begin{table}[h]
\caption{\label{tab:default_parameters}
    Default parameters for generating the synthetic workloads and server. "S1@AGX" means running S1 setting on Jetson AGX Orin.
}
\begin{tabular}{c|ccccccc}
\toprule
Setting & $\gamma$ & k & $\alpha$ & $R$ & $cv$ & $[O_l, O_u]$  & $[I_l, I_u]$ \\ \hline
S1@AGX  & 20 & 3 & 1        & 0.5 & 1    & [8,128] & [8,256]  \\
S2@AGX  & 50 & 3 & 1         &  0.6   &   1   &  [8,128]           &  [8,256]            \\
S3@AGX  & 50 & 3 & 1         &   1  &  1    &  [8,256]           &  [8,256]            \\
S2@Nano & 5 & 3 & 1         &  0.3   &  1    &  [8,128]           &  [8,256]            \\
S3@Nano & 10 & 3 & 1         &  0.6   &  1    &  [8,128]           &  [8,256]            \\
S3@Rasp & 5 & 3 &1         & 0.2    &  1    &  [8,128]           &  [8,128]            \\
\bottomrule
\end{tabular}
\end{table}

\noindent
\textbf{Comparison with \llama.} 
We compare \name\ and \name\ (w/o AAS) with \llama\ for serving multiple LoRA adapters, with results presented in Table~\ref{tab:compare_llamacpp}. Notably, \name\ can serve over 1,000 adapters simultaneously on Jetson Orin AGX, incurring minimal overhead as the number of LoRA models increases. In contrast, \llama\ is limited to serving only 50 adapters due to memory constraints. Overall, \name\ achieves 2-4$\times$ the throughput higher than \llama\, while serving a significantly larger number of adapters across three edge devices.

\begin{table}[h]
\caption{\label{tab:compare_llamacpp}
    Throughput (req/s) comparison between \llama, \name, and \name (w/o AAS) cross devices.
}
\centering    
\scalebox{0.93}{
    \begin{tabular}{c|c|ccc}
    \toprule
    Setting         & n  & \llama & \name & \name (w/o AAS) \\ \hline
    \multirow{4}{*}{S1@AGX} & 20 &    0.11       &        0.45              &   0.45                                        \\
                        & 50 &     0.11      &      0.44                &      0.44                                     \\
                        & 100 &     OOM      &      0.44                &      0.44                                     \\ 
                        & 1000 &     OOM      &         0.42             &       0.44                                    \\ \hline
    \multirow{3}{*}{S2@Nano} & 20 &    0.12       &           0.26           &    0.27                                       \\
                        & 100 &     OOM      &           0.26           &       0.26                                    \\
                        & 500 &      OOM     &            0.25          &        0.26                                   \\ \hline
    \multirow{3}{*}{S3@Rasp} & 20 &     0.05      &         0.19             &       0.20                                    \\
                        & 100 &  OOM         &          0.19            &         0.18                                  \\
                        & 200 &   OOM        &          0.18            &           0.18                                \\ 
    \bottomrule
    \end{tabular}
}
\end{table}

\noindent
\textbf{First token latency and SLO.}
We compare \name\ and \name\ (w/o AAS) with \llama\ in terms of average first token latency and SLO attainment relative to the number of adapters. Table~\ref{tab:slo} shows that \name\ maintains a high SLO across all settings, even when serving a large number of adapters. While the first token latency of \name\ is higher than that of \name (w/o AAS) due to the additional computation for adaptive adapter selection, this does not significantly impact the SLO as shown in Table~\ref{tab:first_token_latency}, ensuring high user satisfaction.
Additionally, we observe that the computational overhead introduced by adaptive adapter selection is roughly equivalent to the time spent on decoding the input prompts.

\begin{table}[h]
\caption{\label{tab:slo}
    SLO comparison between \llama, \name, and \name (w/o AAS) on S3@Nano setting.
}
\begin{tabular}{c|ccc}
\toprule
n    & \llama & \name    &  \name (w/o AAS)  \\ \hline
20   &  1.11\%     & 98.67\% & 100\% \\
100  & OOM   & 98.67\% & 100\% \\
200  & OOM   & 98.67\% & 100\% \\
500  & OOM   & 98.67\% & 100\% \\
1000 & OOM   & 98.00\% & 100\% \\
\bottomrule
\end{tabular}
\end{table}

\begin{table}[h]
\caption{\label{tab:first_token_latency}
    First token latency (s) comparison between \llama, \name, and \name (w/o AAS) on S3@Nano setting.
}
\begin{tabular}{c|ccc}
\toprule
n    & \llama & \name    &  \name (w/o AAS)  \\ \hline
20   &  206.28     & 0.51 & 0.29 \\
100  & OOM   & 0.54 & 0.29 \\
200  & OOM   & 0.54 & 0.32 \\
500  & OOM   & 0.56 & 0.33 \\
1000 & OOM   & 0.58 & 0.35 \\
\bottomrule
\end{tabular}
\end{table}

\noindent
\textbf{Adapter Locality.}
    Real-world workloads typically exhibit temporal locality, where certain adapters are accessed more frequently due to user behavior or popularity differences. Thus, using a low $\alpha$ creates synthetic workloads with high adapter locality~\cite{alpaserve}, effectively simulating realistic scenarios. We compare \name\ with \llama\ across different adapter locality scenarios under the S1@AGX setting with $50$ adapters.
    Table~\ref{tab:alpha_throughput} shows that the throughput of \llama\ is not sensitive to adapter locality, as \llama\ preloads all adapters into memory and only needs to update adapter weights upon switching adapters. Similarly, the throughput of \name\ remains unaffected by adapter locality due to its use of an LRU-based memory cache, which retains frequently accessed adapters and reduces adapter swapping.
    Table~\ref{tab:alpha_latency} indicates that the average request latency for \llama\ decreases slightly under high adapter locality conditions, benefiting from fewer adapter weight switches. The average request latency of \name\ also decreases in high-locality scenarios, as the LRU cache achieves a higher hit rate, further minimizing latency.

\begin{table}[h]
\caption{\label{tab:alpha_throughput}
    Throughput (req/s) comparison between \llama, \name\ on S1@AGX($n=50$) with different adapter locality.
}
\centering
\begin{tabular}{c|cc}
\toprule
$\alpha$ & \llama & \name \\ \hline
0.5  & 0.11 & 0.45 \\
0.75 & 0.11 & 0.45 \\
1    & 0.11 & 0.44 \\
\bottomrule
\end{tabular}
\end{table}

\begin{table}[h]
\caption{\label{tab:alpha_latency}
    Average request latency (s) comparison between \llama, \name\ on S1@AGX($n=50$) with different adapter locality.
}
\centering
\begin{tabular}{c|cc}
\toprule
$\alpha$ & \llama & \name \\ \hline
0.5  & 8.61 & 2.67 \\
0.75 & 8.79 & 2.67 \\
1    & 8.94 & 2.80 \\
\bottomrule
\end{tabular}
\end{table}

\noindent
\textbf{Workload skewness.}
    Users of multi-tenant LLM applications often access services in irregular patterns, causing sudden spikes when multiple users request simultaneously or when a application triggers bursts of usage. High skewness better represents realistic workloads in LLM serving scenarios~\cite{alpaserve}. The intervals between consecutive requests follow a Gamma distribution. When $cv = 1$, the Gamma distribution simplifies to an exponential distribution, indicating moderate burstiness. A $cv > 1$ signifies significantly greater variability and pronounced bursts in request patterns. We compare \name\ with \llama\ across various workload skewness scenarios under the S1@AGX setting with $50$ adapters.
    Table~\ref{tab:cv_throughput} demonstrates that \llama's throughput significantly decreases and its latency increases under high workload skewness. This degradation occurs because \llama\ processes requests sequentially, causing delays when consecutive requests require different adapters. In contrast, \name\ experiences a modest reduction in throughput and increase in latency due to its enhanced capability for parallel request processing. These results indicate that \name\ maintains robust performance even under highly skewed workloads.
    Specifically, when $cv = 2$, some interval between consecutive requests exceeds the request processing time, resulting in similar performance for both \llama\ and \name\ when waiting for request in a long time.

\begin{table}[h]
\caption{\label{tab:cv_throughput}
    Throughput (req/s) comparison between \llama, \name\ on S1@AGX($n=50$) with different workload skewness.
}
\centering
\begin{tabular}{c|cc}
\toprule
$cv$ & \llama & \name \\ \hline
1     & 0.11 & 0.44 \\
1.25  & 0.10 & 0.23 \\
1.5   & 0.09 & 0.11 \\
2     & 0.03 & 0.03 \\
\bottomrule
\end{tabular}
\end{table}

\begin{table}[h]
\caption{\label{tab:cv_latency}
    Average request latency (s) comparison between \llama, \name\ on S1@AGX($n=50$) with different workload skewness.
}
\centering
\begin{tabular}{c|cc}
\toprule
$cv$ & \llama & \name \\ \hline
1     & 8.61 & 2.80 \\
1.25  & 9.13 & 5.27 \\
1.5   & 10.24 & 9.42 \\
2     & 29.25 & 29.54 \\
\bottomrule
\end{tabular}
\end{table}

\noindent
\textbf{Power consumption.}
We compare \name\  with \llama\ in terms of the power consumption across various settings. Power consumption data is collected using jetson-stats~\cite{jetson-stats}, a monitoring tool for system status on Jetson devices. We record power consumption every second and calculate the average across all requests. Table~\ref{tab:power_consumption} shows that \name\ achieves higher energy efficiency than \llama. Moreover, \name\ achieves higher throughput while requiring less computation time to process a given number of requests, resulting in greater power savings.

\begin{table}[h]
\caption{\label{tab:power_consumption}
    Power consumption (Watt) comparison between \llama\ and \name\ cross devices.
}
\begin{tabular}{c|cc}
\toprule
Setting & \llama & \name \\ \hline
S1@AGX ($n$=20)        &  32.16      &   28.04   \\
S2@AGX ($n$=50)       &  24.43      &   24.42    \\
S2@Nano ($n$=20)      &  10.27     &   8.67   \\
\bottomrule
\end{tabular}
\end{table}

\begin{figure*}
    \centering
    \includegraphics[width=\linewidth]{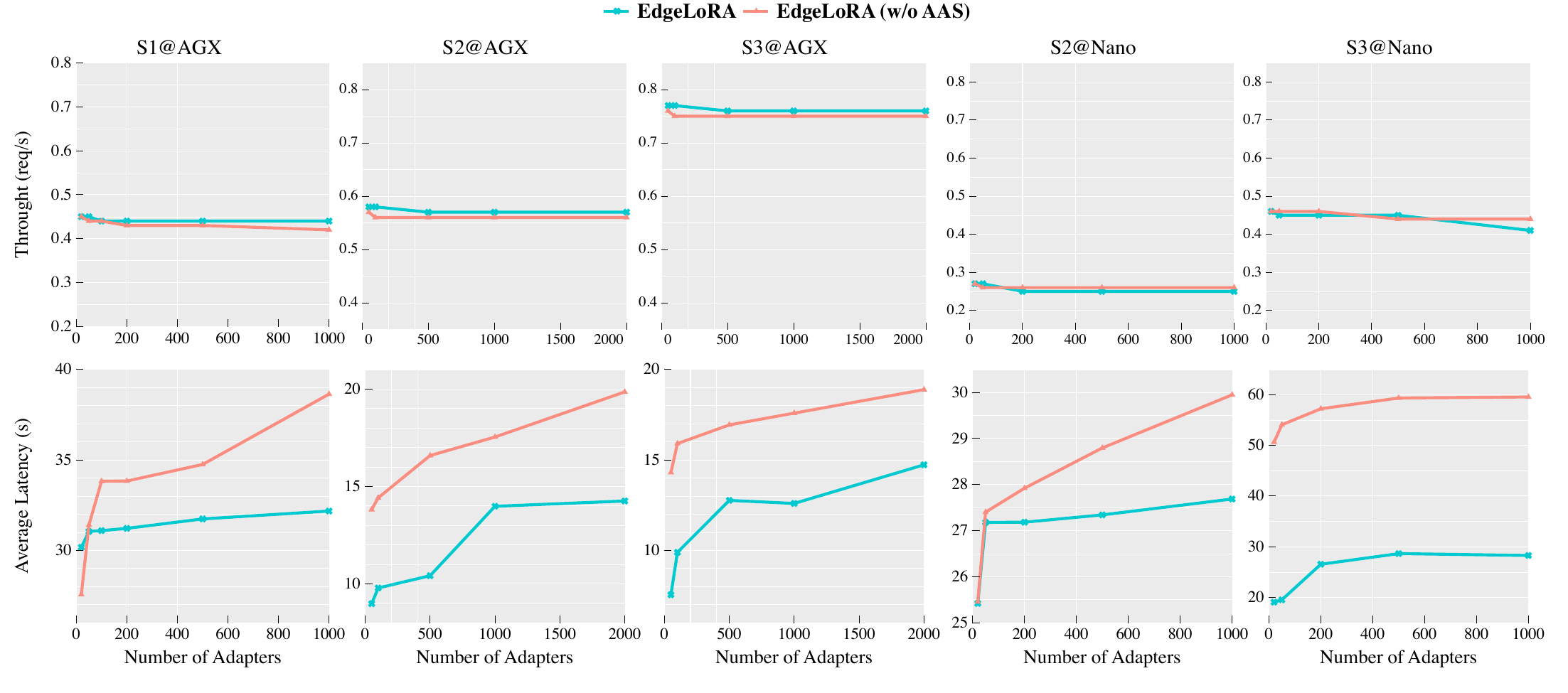}
    \caption{Throughput and average request latency of \name\ and \name\ (w/o AAS) under varying numbers of adapters. Both of them demonstrate scalability to a large number of adapters with similar throughput.}
    \label{fig:comparison_variants}
\end{figure*}

\noindent
\textbf{Comparison with own variants.}
Since \llama\ cannot serve a large number of adapters, our comparison focuses primarily on \name\ and \name\ (w/o AAS). Figure~\ref{fig:comparison_variants} illustrates how both systems scale with the number of adapters. On both Jetson AGX Orin and Jetson Orin Nano, \name\ achieves similar throughput to \name\ (w/o AAS). This latency gap is due to \name\ leverages adaptive adapter selection to maximize the utilization of in-memory adapters.
When more requests can utilize in-memory adapters, computational parallelism increases, allowing multiple requests to be processed simultaneously without waiting for adapter loading operations.
As a result, \name\ exhibits lower latency than \name\ (w/o AAS) on both Jetson AGX Orin and Jetson Orin Nano.
As the number of adapters increases, \name\ maintains stable throughput, and latency increases gradually. However, once the number of adapters exceeds a certain threshold, the latency stabilizes. This stability is due to efficient memory management, where the overhead of swapping adapters remains constant as the number of adapters grows. Consequently, \name\ can scale to handle a significantly large number of adapters without additional overhead, with the only constraint being disk capacity.

\begin{table*}
\caption{\label{tab:adapter_router}
    Adapter router accuracy evaluated with LLaMA3.1-8B-Instruct. 
}
\centering
    \begin{tabular}{l|c c c c c c c}
    \toprule
    \textbf{Model} & \textbf{IFEval} & \textbf{BBH} & \textbf{MATH} & \textbf{GPQA} & \textbf{MMLU-PRO} & \textbf{Average} \\
    \hline
    Llama-3.1-8B-Instruct & 41.84 & 51.22 & 13.82 & 34.95 & 37.85 & 35.94 \\
    Llama-Spark & 43.45 & 52.30  & 13.45 & 31.79 & 38.91 & 35.98  \\
    Defne-llama3.1-8B & 40.92 & 53.10  & 14.56 & 32.42 & 38.82 & 35.96 \\
    Hercules-6.1-Llama-3.1-8B & 47.13 & 51.09 & 13.54 & 32.63 & 37.42 & 36.36 \\
    Llama3.1-8B-ShiningValiant2 & 18.16 & 44.08 & 8.53  & 32.11 & 32.62 & 27.10   \\
    Llama-3.1-8B-German-ORPO & 41.38 & 50.10  & 0.19  & 32.95 & 33.72 & 31.67 \\
    Llama-3.1-SauerkrautLM-8b-Instruct & 45.52 & 51.85 & 15.40  & 33.16 & 39.57 & 37.10   \\
\rowcolor{skyblue}    \textbf{Adapter Router (Our Approach)} & 46.22 & 53.60  & 13.82 & 38.74 & 38.74 & 38.22 \\
    \bottomrule
    \end{tabular}
\end{table*}

\subsection{Adapter Router Performance}

We use five key datasets to generate data for both training and testing. The datasets were selected from the Open LLM Leaderboard~\cite{transformers-library}, as they assess a broad range of reasoning abilities and general knowledge across multiple domains. We randomly split the data, using 80\% for training and the remaining 20\% for testing. Detailed information about these datasets is provided below:

\begin{itemize}
    \item \textbf{IFEval}~\cite{ifeval} A dataset for testing a model's ability to follow explicit instructions like formatting or keyword inclusion. It focuses on adherence to instructions using strict, rigorous metrics.
    
    \item \textbf{BBH}~\cite{bbh} A subset of 23 challenging BigBench tasks using objective metrics to evaluate models. Tasks include multistep reasoning, language understanding, and world knowledge, offering insights into model capabilities.
    
    \item \textbf{MATH}~\cite{math-dataset} A dataset of high-school competition problems formatted in LaTeX and Asymptote. Only level-5 MATH questions are included, requiring specific output formats.
    
    \item \textbf{GPQA}~\cite{gpqa} A knowledge dataset with questions from PhD-level experts in biology, physics, and chemistry. It is highly validated, gated for restricted access, and avoids plain text examples to minimize contamination.
    
    
    \item \textbf{MMLU-PRO}~\cite{mmlu-pro} A refined version of MMLU addressing noisy data and contamination. It increases difficulty with 10-choice questions and expert-reviewed content, providing a higher-quality assessment.
\end{itemize}


We collected six well fine-tuned models based on Llama3.1-8B-instruct~\cite{llama3.1} from the Huggingface Hub: Llama-Spark~\footnote{\url{https://huggingface.co/VAGOsolutions/Llama-3.1-SauerkrautLM-8b-Instruct}}, Defne-llama3.1-8B~\footnote{\url{https://huggingface.co/Eurdem/Defne-llama3.1-8B}}, Hercules-6.1-Llama-3.1-8B~\footnote{\url{https://huggingface.co/Locutusque/Hercules-6.1-Llama-3.1-8B}}, Llama3.1-8B-ShiningVa liant2~\footnote{\url{https://huggingface.co/ValiantLabs/Llama3.1-8B-ShiningValiant2}}, Llama-3.1-8B-German-ORPO~\footnote{\url{https://huggingface.co/Nekochu/Llama-3.1-8B-German-ORPO}}, and Llama-3.1-SauerkrautL M-8b-Instruct~\footnote{\url{https://huggingface.co/VAGOsolutions/Llama-3.1-SauerkrautLM-8b-Instruct}}. The pretrained model is one of the most powerful LLMs nowadays. And each fine-tuned model excels at least in one task, outperforming the pretrained model. We chose to fine-tune the adapter router based on the same pretrained model. 
The model was trained for 3 epochs with a learning rate of $ 1\mathrm{e}{-5} $, using a linear learning rate scheduler and the AdamW optimizer. LoRA-specific parameters included a rank of 32, an alpha value of 64, a dropout rate of 0.05, and targeted layers as the Q, K, V, Down, and Up layers. 
The fine-tuning process took approximately 8 hours using two NVIDIA RTX A6000 Ada Generation GPUs.
After fine-tuning, we evaluated the performance of the adapter router on each task with the test data. Table~\ref{tab:adapter_router} shows that the adapter router consistently outperforms individual adapters by dynamically assigning prompts to the most suitable adapter. 
The performance ceiling of the adapter router is determined by the optimal adapter selection for each prompt, thus inherently constrained by the maximum performance of the individual adapters.

\subsection{Ablation Study}

\subsubsection{DVFS (Dynamic Voltage and Frequency Scaling) of Jetson}

Jetson devices support multiple energy modes, allowing configuration of the number of active cores, their frequencies, and memory frequency to accommodate different power envelopes. Specifically, the Jetson Orin AGX supports TDPs (Thermal Design Power) of 50W, 30W, and 15W, while the Jetson Orin Nano offers 15W and 7W modes. Leveraging this functionality, we conducted experiments to evaluate how these devices perform in executing LLMs under varying power budgets and corresponding performance constraints. 
Table~\ref{tab:ablation_tdp} presents the results of running S1, S2 and S3 on the Jetson Orin AGX under three different TDP levels. The results demonstrate that lower TDP levels constrain the throughput of the serving system.

\begin{table}
\caption{\label{tab:ablation_tdp}
    Throughput (req/s) on Jetson devices under different TDPs.
}
\begin{tabular}{c|ccc}
\toprule
TDP & S1@AGX & S2@AGX & S3@AGX \\ \hline
50W &  0.45  & 0.57  & 0.76 \\
30W &  0.31  &  0.51 & 0.24 \\
15W &  0.13  &  0.22 & 0.14 \\
\bottomrule
\end{tabular}
\end{table}

\subsubsection{Number of slots}

\name\ utilizes a slot state machine to handle concurrent requests. The number of slots can be manually configured to control how many requests are processed simultaneously, while additional requests are queued. A larger number of slots allows more requests to be processed concurrently. When slots are in the processing or generation states, they can be grouped into the same batch, leading to larger batch sizes. 
We conducted experiments to evaluate the impact of the number of slots on serving throughput. Table~\ref{tab:ablation_slot} presents the results for running S1, S2, and S3 on the Jetson Nano Orin with three different slot configurations. The results indicate that a larger number of slots enhances parallel computation capabilities.

\begin{table}[h]
\caption{\label{tab:ablation_slot}
    Throughput (req/s) on Jetson Orin Nano using different number of slots.
}
\begin{tabular}{c|cc}
\toprule
Number of slots & S2@Nano & S3@Nano  \\ \hline
1 &  0.07  &  0.23  \\
5 &  0.17  &  0.39   \\
10 &  0.27  &  0.46  \\
20 &  0.43  &  0.56   \\
\bottomrule
\end{tabular}
\end{table}

\section{Related Work}\label{sec:related_work}


\textbf{LLM Serving}
Numerous studies have sought to optimize the efficiency of LLM serving. vLLM\cite{vllm} introduced a memory-efficient solution based on PagedAttention, which effectively manages key-value cache memory using concepts inspired by classical virtual memory paging. This method significantly reduces memory fragmentation and enhances GPU memory utilization during LLM inference, achieving obviously throughput improvements. However, vLLM can only be deployed on server cluster. In contrast, PowerInfer\cite{powerinfer} focuses on LLM serving using consumer-grade GPUs by implementing a GPU-CPU hybrid inference engine, aiming to enhance performance on personal computers rather than large-scale edge deployments. Both vLLM and PowerInfer do not explicitly address the efficiency challenges of LoRA inference. PipeInfer\cite{pipeinfer} employs speculative execution across distributed server clusters to reduce latency during token generation; however, this architecture is more suited for server-side environments and lacks emphasis on scenarios constrained by edge resources. Llumnix\cite{llumnix} addresses the challenges of scheduling to mitigate severe queuing delays and enhance load balancing in LLM serving, but its primary focus is on dynamic scheduling rather than optimizing efficiency specifically for LoRA adapters. 
InfiniGen~\cite{infinigen} optimizes memory and computation during long-text generation using dynamic key-value cache management, but its reliance on speculative attention computations. 
LServe~\cite{lserve} efficiently accelerates long-sequence LLM serving using hybrid sparse attention. 
MLC-LLM~\cite{mlc-llm} is a machine learning compiler and high-performance deployment engine for LLMs, which support multiple backend of edge devices. But it currently lacks support for serving LoRA models.
Finally, Soter~\cite{soter} offers a secure and efficient partitioning approach to safeguard model confidentiality on edge devices, yet its scope is limited to general neural networks rather than LLMs.

\noindent
\textbf{Efficient LoRA Inference}
The efficient inference of LoRA in LLMs has garnered considerable research attention. S-LoRA\cite{slora} was developed to serve thousands of concurrent LoRA adapters by managing them within unified memory through a unified paging system. Although S-LoRA enhances throughput, it overlooks the issue of high latency caused by unmerged LoRA during inference, which restricts its applicability in latency-sensitive settings. dLoRA\cite{dlora} addresses this limitation by dynamically merging and unmerging adapters and migrating requests between worker replicas, thereby balancing throughput and reducing latency. Nevertheless, dLoRA does not fully resolve the challenge of accelerating unmerged LoRA inference and remains unsuitable for edge device deployment. V-LoRA\cite{vlora} introduces an adaptive-tiling approach for batching LoRA adapters, enabling efficient computation of concurrent heterogeneous LoRA adapters. However, V-LoRA is specifically tailored to vision applications and does not generalize to broader LLM tasks. Punica\cite{punica} provides multi-tenant LoRA serving by batching the pretrained model weights, but it fails to consider batching for the LoRA adapters themselves, thereby limiting further efficiency gains. Moreover, Punica's architecture is designed primarily for shared GPU clusters, presenting challenges for deployment on edge devices. Joint compression~\cite{joint-compression} serves thousands of LoRA adapters efficiently by employing joint compression techniques, enabling scalability and high throughput while maintaining model performance. However, it would sacrifice some fine-tuned model performance, which is critical if precision or task-specific accuracy is crucial.

\noindent
\textbf{Optimize Edge LLM Serving with Algorithm Techniques}
Optimizing the serving of LLMs on edge devices edge devices has attracted significant attention due to inherent computational and memory constraints, prompting various algorithmic optimizations. 
Quantization techniques~\cite{awq, gptq, smoothquant} significantly reduce memory requirements and computational overhead without substantially degrading accuracy. 
Pruning techniques have also been explored to enhance computational efficiency by eliminating redundant model parameters including structured methods~\cite{cofi, llm-pruner}, and unstructured methods~\cite{sparsegpt}. 
Knowledge distillation strategies~\cite{minilm, lion} transfer capabilities from larger models to compact, resource-friendly alternatives, enabling efficient inference. 
Moreover, innovative runtime optimizations such as split inference~\cite{petals, voltage} and speculative decoding~\cite{spectr, voltage} further mitigate latency and bandwidth issues by intelligently distributing workload between cloud and edge components.

\noindent
\textbf{Parameter-efficient Fine-tuning}
Recent advances in PEFT for large pre‑trained language models have demonstrated that updating only a small subset of parameters can yield performance on par with full fine‑tuning. State‑of‑the‑art PEFT techniques include LoRA~\cite{lora}, prefix‑tuning~\cite{prefix-tuning}, P‑Tuning~\cite{p-tuning}, prompt tuning~\cite{prompt-tuning}, AdaLoRA~\cite{adalora}, and IA$^3$~\cite{ia3}. 
More recently, FLoRA~\cite{flora} introduces a federated fine‑tuning framework that leverages heterogeneous LoRA to enable privacy‑preserving, distributed adaptation of LLM. Although our work focuses on LoRA, the same principles readily extend to other PEFT methods.
\section{Conclusion}\label{sec:conclusion}

In this work, we presented \name, an efficient multi-tenant LLM serving system designed for edge devices to address the challenges of serving multiple LoRA adapters. By introducing adaptive adapter selection, heterogeneous memory management, and batched LoRA inference, \name\ eliminates manual adapter selection, optimizes memory usage, and improves computational efficiency. Our system achieves significant improvements in throughput, scalability, and energy efficiency, demonstrating its ability to manage dynamic workloads across multi-tenant edge environments. Experimental results confirm that \name\ outperforms existing solutions, supporting a larger number of adapters while maintaining low latency and high user satisfaction. These findings highlight the potential of \name\ to enable advanced, multi-tenant LLM applications on edge devices.

\bibliographystyle{ACM-Reference-Format}
\bibliography{reference}


\begin{thebibliography}{100}


\ifx \showCODEN    \undefined \def \showCODEN     #1{\unskip}     \fi
\ifx \showDOI      \undefined \def \showDOI       #1{#1}\fi
\ifx \showISBNx    \undefined \def \showISBNx     #1{\unskip}     \fi
\ifx \showISBNxiii \undefined \def \showISBNxiii  #1{\unskip}     \fi
\ifx \showISSN     \undefined \def \showISSN      #1{\unskip}     \fi
\ifx \showLCCN     \undefined \def \showLCCN      #1{\unskip}     \fi
\ifx \shownote     \undefined \def \shownote      #1{#1}          \fi
\ifx \showarticletitle \undefined \def \showarticletitle #1{#1}   \fi
\ifx \showURL      \undefined \def \showURL       {\relax}        \fi
\providecommand\bibfield[2]{#2}
\providecommand\bibinfo[2]{#2}
\providecommand\natexlab[1]{#1}
\providecommand\showeprint[2][]{arXiv:#2}

\bibitem[Abdin et~al\mbox{.}(2024)]%
        {phi3}
\bibfield{author}{\bibinfo{person}{Marah Abdin}, \bibinfo{person}{Jyoti Aneja},
  \bibinfo{person}{Hany Awadalla}, \bibinfo{person}{Ahmed Awadallah},
  \bibinfo{person}{Ammar~Ahmad Awan}, \bibinfo{person}{Nguyen Bach},
  \bibinfo{person}{Amit Bahree}, \bibinfo{person}{Arash Bakhtiari},
  \bibinfo{person}{Jianmin Bao}, \bibinfo{person}{Harkirat Behl},
  \bibinfo{person}{Alon Benhaim}, \bibinfo{person}{Misha Bilenko},
  \bibinfo{person}{Johan Bjorck}, \bibinfo{person}{Sébastien Bubeck},
  \bibinfo{person}{Martin Cai}, \bibinfo{person}{Qin Cai},
  \bibinfo{person}{Vishrav Chaudhary}, \bibinfo{person}{Dong Chen},
  \bibinfo{person}{Dongdong Chen}, \bibinfo{person}{Weizhu Chen},
  \bibinfo{person}{Yen-Chun Chen}, \bibinfo{person}{Yi-Ling Chen},
  \bibinfo{person}{Hao Cheng}, \bibinfo{person}{Parul Chopra},
  \bibinfo{person}{Xiyang Dai}, \bibinfo{person}{Matthew Dixon},
  \bibinfo{person}{Ronen Eldan}, \bibinfo{person}{Victor Fragoso},
  \bibinfo{person}{Jianfeng Gao}, \bibinfo{person}{Mei Gao},
  \bibinfo{person}{Min Gao}, \bibinfo{person}{Amit Garg},
  \bibinfo{person}{Allie~Del Giorno}, \bibinfo{person}{Abhishek Goswami},
  \bibinfo{person}{Suriya Gunasekar}, \bibinfo{person}{Emman Haider},
  \bibinfo{person}{Junheng Hao}, \bibinfo{person}{Russell~J. Hewett},
  \bibinfo{person}{Wenxiang Hu}, \bibinfo{person}{Jamie Huynh},
  \bibinfo{person}{Dan Iter}, \bibinfo{person}{Sam~Ade Jacobs},
  \bibinfo{person}{Mojan Javaheripi}, \bibinfo{person}{Xin Jin},
  \bibinfo{person}{Nikos Karampatziakis}, \bibinfo{person}{Piero Kauffmann},
  \bibinfo{person}{Mahoud Khademi}, \bibinfo{person}{Dongwoo Kim},
  \bibinfo{person}{Young~Jin Kim}, \bibinfo{person}{Lev Kurilenko},
  \bibinfo{person}{James~R. Lee}, \bibinfo{person}{Yin~Tat Lee},
  \bibinfo{person}{Yuanzhi Li}, \bibinfo{person}{Yunsheng Li},
  \bibinfo{person}{Chen Liang}, \bibinfo{person}{Lars Liden},
  \bibinfo{person}{Xihui Lin}, \bibinfo{person}{Zeqi Lin}, \bibinfo{person}{Ce
  Liu}, \bibinfo{person}{Liyuan Liu}, \bibinfo{person}{Mengchen Liu},
  \bibinfo{person}{Weishung Liu}, \bibinfo{person}{Xiaodong Liu},
  \bibinfo{person}{Chong Luo}, \bibinfo{person}{Piyush Madan},
  \bibinfo{person}{Ali Mahmoudzadeh}, \bibinfo{person}{David Majercak},
  \bibinfo{person}{Matt Mazzola}, \bibinfo{person}{Caio César~Teodoro Mendes},
  \bibinfo{person}{Arindam Mitra}, \bibinfo{person}{Hardik Modi},
  \bibinfo{person}{Anh Nguyen}, \bibinfo{person}{Brandon Norick},
  \bibinfo{person}{Barun Patra}, \bibinfo{person}{Daniel Perez-Becker},
  \bibinfo{person}{Thomas Portet}, \bibinfo{person}{Reid Pryzant},
  \bibinfo{person}{Heyang Qin}, \bibinfo{person}{Marko Radmilac},
  \bibinfo{person}{Liliang Ren}, \bibinfo{person}{Gustavo de Rosa},
  \bibinfo{person}{Corby Rosset}, \bibinfo{person}{Sambudha Roy},
  \bibinfo{person}{Olatunji Ruwase}, \bibinfo{person}{Olli Saarikivi},
  \bibinfo{person}{Amin Saied}, \bibinfo{person}{Adil Salim},
  \bibinfo{person}{Michael Santacroce}, \bibinfo{person}{Shital Shah},
  \bibinfo{person}{Ning Shang}, \bibinfo{person}{Hiteshi Sharma},
  \bibinfo{person}{Yelong Shen}, \bibinfo{person}{Swadheen Shukla},
  \bibinfo{person}{Xia Song}, \bibinfo{person}{Masahiro Tanaka},
  \bibinfo{person}{Andrea Tupini}, \bibinfo{person}{Praneetha Vaddamanu},
  \bibinfo{person}{Chunyu Wang}, \bibinfo{person}{Guanhua Wang},
  \bibinfo{person}{Lijuan Wang}, \bibinfo{person}{Shuohang Wang},
  \bibinfo{person}{Xin Wang}, \bibinfo{person}{Yu Wang},
  \bibinfo{person}{Rachel Ward}, \bibinfo{person}{Wen Wen},
  \bibinfo{person}{Philipp Witte}, \bibinfo{person}{Haiping Wu},
  \bibinfo{person}{Xiaoxia Wu}, \bibinfo{person}{Michael Wyatt},
  \bibinfo{person}{Bin Xiao}, \bibinfo{person}{Can Xu},
  \bibinfo{person}{Jiahang Xu}, \bibinfo{person}{Weijian Xu},
  \bibinfo{person}{Jilong Xue}, \bibinfo{person}{Sonali Yadav},
  \bibinfo{person}{Fan Yang}, \bibinfo{person}{Jianwei Yang},
  \bibinfo{person}{Yifan Yang}, \bibinfo{person}{Ziyi Yang},
  \bibinfo{person}{Donghan Yu}, \bibinfo{person}{Lu Yuan},
  \bibinfo{person}{Chenruidong Zhang}, \bibinfo{person}{Cyril Zhang},
  \bibinfo{person}{Jianwen Zhang}, \bibinfo{person}{Li~Lyna Zhang},
  \bibinfo{person}{Yi Zhang}, \bibinfo{person}{Yue Zhang},
  \bibinfo{person}{Yunan Zhang}, {and} \bibinfo{person}{Xiren Zhou}.}
  \bibinfo{year}{2024}\natexlab{}.
\newblock \bibinfo{title}{Phi-3 Technical Report: A Highly Capable Language
  Model Locally on Your Phone}.
\newblock
\showeprint[arxiv]{2404.14219}~[cs.CL]
\urldef\tempurl%
\url{https://arxiv.org/abs/2404.14219}
\showURL{%
\tempurl}


\bibitem[Adiwardana et~al\mbox{.}(2020)]%
        {chatbot}
\bibfield{author}{\bibinfo{person}{Daniel Adiwardana},
  \bibinfo{person}{Minh-Thang Luong}, \bibinfo{person}{David~R So},
  \bibinfo{person}{Jamie Hall}, \bibinfo{person}{Noah Fiedel},
  \bibinfo{person}{Romal Thoppilan}, \bibinfo{person}{Zi Yang},
  \bibinfo{person}{Apoorv Kulshreshtha}, \bibinfo{person}{Gaurav Nemade},
  \bibinfo{person}{Yifeng Lu}, {et~al\mbox{.}}}
  \bibinfo{year}{2020}\natexlab{}.
\newblock \showarticletitle{Towards a human-like open-domain chatbot}.
\newblock \bibinfo{journal}{\emph{arXiv preprint arXiv:2001.09977}}
  (\bibinfo{year}{2020}).
\newblock


\bibitem[Anthropic(2023)]%
        {claude}
\bibfield{author}{\bibinfo{person}{Anthropic}.}
  \bibinfo{year}{2023}\natexlab{}.
\newblock \bibinfo{title}{Claude 3 Model Card}.
\newblock
\urldef\tempurl%
\url{https://www.anthropic.com/model-card-claude-3}
\showURL{%
\tempurl}
\newblock
\shownote{Accessed: 2024-12-03}.


\bibitem[Bai et~al\mbox{.}(2023)]%
        {qwen}
\bibfield{author}{\bibinfo{person}{Jinze Bai}, \bibinfo{person}{Shuai Bai},
  \bibinfo{person}{Yunfei Chu}, \bibinfo{person}{Zeyu Cui},
  \bibinfo{person}{Kai Dang}, \bibinfo{person}{Xiaodong Deng},
  \bibinfo{person}{Yang Fan}, \bibinfo{person}{Wenbin Ge}, \bibinfo{person}{Yu
  Han}, \bibinfo{person}{Fei Huang}, \bibinfo{person}{Binyuan Hui},
  \bibinfo{person}{Luo Ji}, \bibinfo{person}{Mei Li}, \bibinfo{person}{Junyang
  Lin}, \bibinfo{person}{Runji Lin}, \bibinfo{person}{Dayiheng Liu},
  \bibinfo{person}{Gao Liu}, \bibinfo{person}{Chengqiang Lu},
  \bibinfo{person}{Keming Lu}, \bibinfo{person}{Jianxin Ma},
  \bibinfo{person}{Rui Men}, \bibinfo{person}{Xingzhang Ren},
  \bibinfo{person}{Xuancheng Ren}, \bibinfo{person}{Chuanqi Tan},
  \bibinfo{person}{Sinan Tan}, \bibinfo{person}{Jianhong Tu},
  \bibinfo{person}{Peng Wang}, \bibinfo{person}{Shijie Wang},
  \bibinfo{person}{Wei Wang}, \bibinfo{person}{Shengguang Wu},
  \bibinfo{person}{Benfeng Xu}, \bibinfo{person}{Jin Xu}, \bibinfo{person}{An
  Yang}, \bibinfo{person}{Hao Yang}, \bibinfo{person}{Jian Yang},
  \bibinfo{person}{Shusheng Yang}, \bibinfo{person}{Yang Yao},
  \bibinfo{person}{Bowen Yu}, \bibinfo{person}{Hongyi Yuan},
  \bibinfo{person}{Zheng Yuan}, \bibinfo{person}{Jianwei Zhang},
  \bibinfo{person}{Xingxuan Zhang}, \bibinfo{person}{Yichang Zhang},
  \bibinfo{person}{Zhenru Zhang}, \bibinfo{person}{Chang Zhou},
  \bibinfo{person}{Jingren Zhou}, \bibinfo{person}{Xiaohuan Zhou}, {and}
  \bibinfo{person}{Tianhang Zhu}.} \bibinfo{year}{2023}\natexlab{}.
\newblock \bibinfo{title}{Qwen Technical Report}.
\newblock
\showeprint[arxiv]{2309.16609}~[cs.CL]
\urldef\tempurl%
\url{https://arxiv.org/abs/2309.16609}
\showURL{%
\tempurl}


\bibitem[Biderman et~al\mbox{.}(2024)]%
        {lora-forget}
\bibfield{author}{\bibinfo{person}{Dan Biderman}, \bibinfo{person}{Jacob
  Portes}, \bibinfo{person}{Jose Javier~Gonzalez Ortiz},
  \bibinfo{person}{Mansheej Paul}, \bibinfo{person}{Philip Greengard},
  \bibinfo{person}{Connor Jennings}, \bibinfo{person}{Daniel King},
  \bibinfo{person}{Sam Havens}, \bibinfo{person}{Vitaliy Chiley},
  \bibinfo{person}{Jonathan Frankle}, \bibinfo{person}{Cody Blakeney}, {and}
  \bibinfo{person}{John~P. Cunningham}.} \bibinfo{year}{2024}\natexlab{}.
\newblock \bibinfo{title}{LoRA Learns Less and Forgets Less}.
\newblock
\showeprint[arxiv]{2405.09673}~[cs.LG]
\urldef\tempurl%
\url{https://arxiv.org/abs/2405.09673}
\showURL{%
\tempurl}


\bibitem[Bonghi(2023)]%
        {jetson-stats}
\bibfield{author}{\bibinfo{person}{Raffaello Bonghi}.}
  \bibinfo{year}{2023}\natexlab{}.
\newblock \bibinfo{title}{Jetson-Stats}.
\newblock
  \bibinfo{howpublished}{\url{https://github.com/rbonghi/jetson_stats}}.
\newblock
\newblock
\shownote{Accessed: 2024-12-07}.


\bibitem[Borzunov et~al\mbox{.}(2023)]%
        {petals}
\bibfield{author}{\bibinfo{person}{Alexander Borzunov}, \bibinfo{person}{Max
  Ryabinin}, \bibinfo{person}{Artem Chumachenko}, \bibinfo{person}{Dmitry
  Baranchuk}, \bibinfo{person}{Tim Dettmers}, \bibinfo{person}{Younes Belkada},
  \bibinfo{person}{Pavel Samygin}, {and} \bibinfo{person}{Colin~A Raffel}.}
  \bibinfo{year}{2023}\natexlab{}.
\newblock \showarticletitle{Distributed inference and fine-tuning of large
  language models over the internet}.
\newblock \bibinfo{journal}{\emph{Advances in neural information processing
  systems}}  \bibinfo{volume}{36} (\bibinfo{year}{2023}),
  \bibinfo{pages}{12312--12331}.
\newblock


\bibitem[Brown et~al\mbox{.}(1990)]%
        {machine-translation}
\bibfield{author}{\bibinfo{person}{Peter~F Brown}, \bibinfo{person}{John
  Cocke}, \bibinfo{person}{Stephen~A Della~Pietra}, \bibinfo{person}{Vincent~J
  Della~Pietra}, \bibinfo{person}{Frederick Jelinek}, \bibinfo{person}{John
  Lafferty}, \bibinfo{person}{Robert~L Mercer}, {and} \bibinfo{person}{Paul~S
  Roossin}.} \bibinfo{year}{1990}\natexlab{}.
\newblock \showarticletitle{A statistical approach to machine translation}.
\newblock \bibinfo{journal}{\emph{Computational linguistics}}
  \bibinfo{volume}{16}, \bibinfo{number}{2} (\bibinfo{year}{1990}),
  \bibinfo{pages}{79--85}.
\newblock


\bibitem[Brown et~al\mbox{.}(2020)]%
        {gpt3}
\bibfield{author}{\bibinfo{person}{Tom~B. Brown}, \bibinfo{person}{Benjamin
  Mann}, \bibinfo{person}{Nick Ryder}, \bibinfo{person}{Melanie Subbiah},
  \bibinfo{person}{Jared Kaplan}, \bibinfo{person}{Prafulla Dhariwal},
  \bibinfo{person}{Arvind Neelakantan}, \bibinfo{person}{Pranav Shyam},
  \bibinfo{person}{Girish Sastry}, \bibinfo{person}{Amanda Askell},
  \bibinfo{person}{Sandhini Agarwal}, \bibinfo{person}{Ariel Herbert-Voss},
  \bibinfo{person}{Gretchen Krueger}, \bibinfo{person}{Tom Henighan},
  \bibinfo{person}{Rewon Child}, \bibinfo{person}{Aditya Ramesh},
  \bibinfo{person}{Daniel~M. Ziegler}, \bibinfo{person}{Jeffrey Wu},
  \bibinfo{person}{Clemens Winter}, \bibinfo{person}{Christopher Hesse},
  \bibinfo{person}{Mark Chen}, \bibinfo{person}{Eric Sigler},
  \bibinfo{person}{Mateusz Litwin}, \bibinfo{person}{Scott Gray},
  \bibinfo{person}{Benjamin Chess}, \bibinfo{person}{Jack Clark},
  \bibinfo{person}{Christopher Berner}, \bibinfo{person}{Sam McCandlish},
  \bibinfo{person}{Alec Radford}, \bibinfo{person}{Ilya Sutskever}, {and}
  \bibinfo{person}{Dario Amodei}.} \bibinfo{year}{2020}\natexlab{}.
\newblock \bibinfo{title}{Language Models are Few-Shot Learners}.
\newblock
\showeprint[arxiv]{2005.14165}~[cs.CL]
\urldef\tempurl%
\url{https://arxiv.org/abs/2005.14165}
\showURL{%
\tempurl}


\bibitem[Brüel-Gabrielsson et~al\mbox{.}(2024)]%
        {joint-compression}
\bibfield{author}{\bibinfo{person}{Rickard Brüel-Gabrielsson},
  \bibinfo{person}{Jiacheng Zhu}, \bibinfo{person}{Onkar Bhardwaj},
  \bibinfo{person}{Leshem Choshen}, \bibinfo{person}{Kristjan Greenewald},
  \bibinfo{person}{Mikhail Yurochkin}, {and} \bibinfo{person}{Justin Solomon}.}
  \bibinfo{year}{2024}\natexlab{}.
\newblock \bibinfo{title}{Compress then Serve: Serving Thousands of LoRA
  Adapters with Little Overhead}.
\newblock
\showeprint[arxiv]{2407.00066}~[cs.DC]
\urldef\tempurl%
\url{https://arxiv.org/abs/2407.00066}
\showURL{%
\tempurl}


\bibitem[Butler et~al\mbox{.}(2024)]%
        {pipeinfer}
\bibfield{author}{\bibinfo{person}{Branden Butler}, \bibinfo{person}{Sixing
  Yu}, \bibinfo{person}{Arya Mazaheri}, {and} \bibinfo{person}{Ali Jannesari}.}
  \bibinfo{year}{2024}\natexlab{}.
\newblock \showarticletitle{PipeInfer: Accelerating LLM Inference using
  Asynchronous Pipelined Speculation}.
\newblock \bibinfo{journal}{\emph{arXiv preprint arXiv:2407.11798}}
  (\bibinfo{year}{2024}).
\newblock


\bibitem[Chavan et~al\mbox{.}(2023)]%
        {glora}
\bibfield{author}{\bibinfo{person}{Arnav Chavan}, \bibinfo{person}{Zhuang Liu},
  \bibinfo{person}{Deepak Gupta}, \bibinfo{person}{Eric Xing}, {and}
  \bibinfo{person}{Zhiqiang Shen}.} \bibinfo{year}{2023}\natexlab{}.
\newblock \showarticletitle{One-for-all: Generalized lora for
  parameter-efficient fine-tuning}.
\newblock \bibinfo{journal}{\emph{arXiv preprint arXiv:2306.07967}}
  (\bibinfo{year}{2023}).
\newblock


\bibitem[Chen et~al\mbox{.}(2023b)]%
        {punica}
\bibfield{author}{\bibinfo{person}{Lequn Chen}, \bibinfo{person}{Zihao Ye},
  \bibinfo{person}{Yongji Wu}, \bibinfo{person}{Danyang Zhuo},
  \bibinfo{person}{Luis Ceze}, {and} \bibinfo{person}{Arvind Krishnamurthy}.}
  \bibinfo{year}{2023}\natexlab{b}.
\newblock \bibinfo{title}{Punica: Multi-Tenant LoRA Serving}.
\newblock
\showeprint[arxiv]{2310.18547}~[cs.DC]
\urldef\tempurl%
\url{https://arxiv.org/abs/2310.18547}
\showURL{%
\tempurl}


\bibitem[Chen et~al\mbox{.}(2023a)]%
        {longlora}
\bibfield{author}{\bibinfo{person}{Yukang Chen}, \bibinfo{person}{Shengju
  Qian}, \bibinfo{person}{Haotian Tang}, \bibinfo{person}{Xin Lai},
  \bibinfo{person}{Zhijian Liu}, \bibinfo{person}{Song Han}, {and}
  \bibinfo{person}{Jiaya Jia}.} \bibinfo{year}{2023}\natexlab{a}.
\newblock \showarticletitle{Longlora: Efficient fine-tuning of long-context
  large language models}.
\newblock \bibinfo{journal}{\emph{arXiv preprint arXiv:2309.12307}}
  (\bibinfo{year}{2023}).
\newblock


\bibitem[Chowdhery et~al\mbox{.}(2023)]%
        {palm}
\bibfield{author}{\bibinfo{person}{Aakanksha Chowdhery},
  \bibinfo{person}{Sharan Narang}, \bibinfo{person}{Jacob Devlin},
  \bibinfo{person}{Maarten Bosma}, \bibinfo{person}{Gaurav Mishra},
  \bibinfo{person}{Adam Roberts}, \bibinfo{person}{Paul Barham},
  \bibinfo{person}{Hyung~Won Chung}, \bibinfo{person}{Charles Sutton},
  \bibinfo{person}{Sebastian Gehrmann}, {et~al\mbox{.}}}
  \bibinfo{year}{2023}\natexlab{}.
\newblock \showarticletitle{Palm: Scaling language modeling with pathways}.
\newblock \bibinfo{journal}{\emph{Journal of Machine Learning Research}}
  \bibinfo{volume}{24}, \bibinfo{number}{240} (\bibinfo{year}{2023}),
  \bibinfo{pages}{1--113}.
\newblock


\bibitem[Dettmers et~al\mbox{.}(2024)]%
        {qlora}
\bibfield{author}{\bibinfo{person}{Tim Dettmers}, \bibinfo{person}{Artidoro
  Pagnoni}, \bibinfo{person}{Ari Holtzman}, {and} \bibinfo{person}{Luke
  Zettlemoyer}.} \bibinfo{year}{2024}\natexlab{}.
\newblock \showarticletitle{Qlora: Efficient finetuning of quantized llms}.
\newblock \bibinfo{journal}{\emph{Advances in Neural Information Processing
  Systems}}  \bibinfo{volume}{36} (\bibinfo{year}{2024}).
\newblock


\bibitem[Dosovitskiy et~al\mbox{.}(2021)]%
        {vit}
\bibfield{author}{\bibinfo{person}{Alexey Dosovitskiy}, \bibinfo{person}{Lucas
  Beyer}, \bibinfo{person}{Alexander Kolesnikov}, \bibinfo{person}{Dirk
  Weissenborn}, \bibinfo{person}{Xiaohua Zhai}, \bibinfo{person}{Thomas
  Unterthiner}, \bibinfo{person}{Mostafa Dehghani}, \bibinfo{person}{Matthias
  Minderer}, \bibinfo{person}{Georg Heigold}, \bibinfo{person}{Sylvain Gelly},
  \bibinfo{person}{Jakob Uszkoreit}, {and} \bibinfo{person}{Neil Houlsby}.}
  \bibinfo{year}{2021}\natexlab{}.
\newblock \bibinfo{title}{An Image is Worth 16x16 Words: Transformers for Image
  Recognition at Scale}.
\newblock
\showeprint[arxiv]{2010.11929}~[cs.CV]
\urldef\tempurl%
\url{https://arxiv.org/abs/2010.11929}
\showURL{%
\tempurl}


\bibitem[Dwivedi-Yu et~al\mbox{.}(2022)]%
        {editeval-dataset}
\bibfield{author}{\bibinfo{person}{Jane Dwivedi-Yu}, \bibinfo{person}{Timo
  Schick}, \bibinfo{person}{Zhengbao Jiang}, \bibinfo{person}{Maria Lomeli},
  \bibinfo{person}{Patrick Lewis}, \bibinfo{person}{Gautier Izacard},
  \bibinfo{person}{Edouard Grave}, \bibinfo{person}{Sebastian Riedel}, {and}
  \bibinfo{person}{Fabio Petroni}.} \bibinfo{year}{2022}\natexlab{}.
\newblock \bibinfo{title}{EditEval: An Instruction-Based Benchmark for Text
  Improvements}.
\newblock
\showeprint[arxiv]{2209.13331}~[cs.CL]
\urldef\tempurl%
\url{https://arxiv.org/abs/2209.13331}
\showURL{%
\tempurl}


\bibitem[El-Kassas et~al\mbox{.}(2021)]%
        {summarization-survey}
\bibfield{author}{\bibinfo{person}{Wafaa~S El-Kassas},
  \bibinfo{person}{Cherif~R Salama}, \bibinfo{person}{Ahmed~A Rafea}, {and}
  \bibinfo{person}{Hoda~K Mohamed}.} \bibinfo{year}{2021}\natexlab{}.
\newblock \showarticletitle{Automatic text summarization: A comprehensive
  survey}.
\newblock \bibinfo{journal}{\emph{Expert systems with applications}}
  \bibinfo{volume}{165} (\bibinfo{year}{2021}), \bibinfo{pages}{113679}.
\newblock


\bibitem[Face(2023)]%
        {textgeninference}
\bibfield{author}{\bibinfo{person}{Hugging Face}.}
  \bibinfo{year}{2023}\natexlab{}.
\newblock \bibinfo{title}{Text Generation Inference: Large Language Model Text
  Generation Inference}.
\newblock
  \bibinfo{howpublished}{\url{https://github.com/huggingface/text-generation-inference}}.
\newblock


\bibitem[Frantar and Alistarh(2023)]%
        {sparsegpt}
\bibfield{author}{\bibinfo{person}{Elias Frantar} {and} \bibinfo{person}{Dan
  Alistarh}.} \bibinfo{year}{2023}\natexlab{}.
\newblock \showarticletitle{Sparsegpt: Massive language models can be
  accurately pruned in one-shot}. In \bibinfo{booktitle}{\emph{International
  Conference on Machine Learning}}. PMLR, \bibinfo{pages}{10323--10337}.
\newblock


\bibitem[Frantar et~al\mbox{.}(2022)]%
        {gptq}
\bibfield{author}{\bibinfo{person}{Elias Frantar}, \bibinfo{person}{Saleh
  Ashkboos}, \bibinfo{person}{Torsten Hoefler}, {and} \bibinfo{person}{Dan
  Alistarh}.} \bibinfo{year}{2022}\natexlab{}.
\newblock \showarticletitle{Gptq: Accurate post-training quantization for
  generative pre-trained transformers}.
\newblock \bibinfo{journal}{\emph{arXiv preprint arXiv:2210.17323}}
  (\bibinfo{year}{2022}).
\newblock


\bibitem[Gao et~al\mbox{.}(2023)]%
        {lm-eval}
\bibfield{author}{\bibinfo{person}{Leo Gao}, \bibinfo{person}{Jonathan Tow},
  \bibinfo{person}{Baber Abbasi}, \bibinfo{person}{Stella Biderman},
  \bibinfo{person}{Sid Black}, \bibinfo{person}{Anthony DiPofi},
  \bibinfo{person}{Charles Foster}, \bibinfo{person}{Laurence Golding},
  \bibinfo{person}{Jeffrey Hsu}, \bibinfo{person}{Alain Le~Noac'h},
  \bibinfo{person}{Haonan Li}, \bibinfo{person}{Kyle McDonell},
  \bibinfo{person}{Niklas Muennighoff}, \bibinfo{person}{Chris Ociepa},
  \bibinfo{person}{Jason Phang}, \bibinfo{person}{Laria Reynolds},
  \bibinfo{person}{Hailey Schoelkopf}, \bibinfo{person}{Aviya Skowron},
  \bibinfo{person}{Lintang Sutawika}, \bibinfo{person}{Eric Tang},
  \bibinfo{person}{Anish Thite}, \bibinfo{person}{Ben Wang},
  \bibinfo{person}{Kevin Wang}, {and} \bibinfo{person}{Andy Zou}.}
  \bibinfo{year}{2023}\natexlab{}.
\newblock \bibinfo{title}{A framework for few-shot language model evaluation}.
\newblock
\urldef\tempurl%
\url{https://doi.org/10.5281/zenodo.10256836}
\showDOI{\tempurl}


\bibitem[Gerganov(2023)]%
        {llamacpp}
\bibfield{author}{\bibinfo{person}{Georgi Gerganov}.}
  \bibinfo{year}{2023}\natexlab{}.
\newblock \bibinfo{title}{llama.cpp: LLM inference in C/C++}.
\newblock \bibinfo{howpublished}{\url{https://github.com/ggerganov/llama.cpp}}.
\newblock


\bibitem[Gerganov(2024)]%
        {ggml}
\bibfield{author}{\bibinfo{person}{Georgi Gerganov}.}
  \bibinfo{year}{2024}\natexlab{}.
\newblock \bibinfo{title}{ggml}.
\newblock \bibinfo{howpublished}{\url{https://github.com/ggerganov/ggml}}.
\newblock
\newblock
\shownote{Accessed: 2024-12-03}.


\bibitem[Gong et~al\mbox{.}(2023)]%
        {multimodalgpt}
\bibfield{author}{\bibinfo{person}{Tao Gong}, \bibinfo{person}{Chengqi Lyu},
  \bibinfo{person}{Shilong Zhang}, \bibinfo{person}{Yudong Wang},
  \bibinfo{person}{Miao Zheng}, \bibinfo{person}{Qian Zhao},
  \bibinfo{person}{Kuikun Liu}, \bibinfo{person}{Wenwei Zhang},
  \bibinfo{person}{Ping Luo}, {and} \bibinfo{person}{Kai Chen}.}
  \bibinfo{year}{2023}\natexlab{}.
\newblock \showarticletitle{Multimodal-gpt: A vision and language model for
  dialogue with humans}.
\newblock \bibinfo{journal}{\emph{arXiv preprint arXiv:2305.04790}}
  (\bibinfo{year}{2023}).
\newblock


\bibitem[Grattafiori et~al\mbox{.}(2024)]%
        {llama3.1}
\bibfield{author}{\bibinfo{person}{Aaron Grattafiori},
  \bibinfo{person}{Abhimanyu Dubey}, \bibinfo{person}{Abhinav Jauhri},
  \bibinfo{person}{Abhinav Pandey}, \bibinfo{person}{Abhishek Kadian},
  \bibinfo{person}{Ahmad Al-Dahle}, \bibinfo{person}{Aiesha Letman}, {and}
  \bibinfo{person}{Others}.} \bibinfo{year}{2024}\natexlab{}.
\newblock \bibinfo{title}{The Llama 3 Herd of Models}.
\newblock
\showeprint[arxiv]{2407.21783}~[cs.AI]
\urldef\tempurl%
\url{https://arxiv.org/abs/2407.21783}
\showURL{%
\tempurl}


\bibitem[Graves and Graves(2012)]%
        {lstm}
\bibfield{author}{\bibinfo{person}{Alex Graves} {and} \bibinfo{person}{Alex
  Graves}.} \bibinfo{year}{2012}\natexlab{}.
\newblock \showarticletitle{Long short-term memory}.
\newblock \bibinfo{journal}{\emph{Supervised sequence labelling with recurrent
  neural networks}} (\bibinfo{year}{2012}), \bibinfo{pages}{37--45}.
\newblock


\bibitem[Gupta et~al\mbox{.}(2022)]%
        {instructdial}
\bibfield{author}{\bibinfo{person}{Prakhar Gupta}, \bibinfo{person}{Cathy
  Jiao}, \bibinfo{person}{Yi-Ting Yeh}, \bibinfo{person}{Shikib Mehri},
  \bibinfo{person}{Maxine Eskenazi}, {and} \bibinfo{person}{Jeffrey~P Bigham}.}
  \bibinfo{year}{2022}\natexlab{}.
\newblock \showarticletitle{InstructDial: Improving zero and few-shot
  generalization in dialogue through instruction tuning}.
\newblock \bibinfo{journal}{\emph{arXiv preprint arXiv:2205.12673}}
  (\bibinfo{year}{2022}).
\newblock


\bibitem[Hendrycks et~al\mbox{.}(2021)]%
        {math-dataset}
\bibfield{author}{\bibinfo{person}{Dan Hendrycks}, \bibinfo{person}{Collin
  Burns}, \bibinfo{person}{Saurav Kadavath}, \bibinfo{person}{Akul Arora},
  \bibinfo{person}{Steven Basart}, \bibinfo{person}{Eric Tang},
  \bibinfo{person}{Dawn Song}, {and} \bibinfo{person}{Jacob Steinhardt}.}
  \bibinfo{year}{2021}\natexlab{}.
\newblock \bibinfo{title}{Measuring Mathematical Problem Solving With the MATH
  Dataset}.
\newblock
\showeprint[arxiv]{2103.03874}~[cs.LG]
\urldef\tempurl%
\url{https://arxiv.org/abs/2103.03874}
\showURL{%
\tempurl}


\bibitem[Houlsby et~al\mbox{.}(2019)]%
        {adapter-peft}
\bibfield{author}{\bibinfo{person}{Neil Houlsby}, \bibinfo{person}{Andrei
  Giurgiu}, \bibinfo{person}{Stanislaw Jastrzebski}, \bibinfo{person}{Bruna
  Morrone}, \bibinfo{person}{Quentin De~Laroussilhe}, \bibinfo{person}{Andrea
  Gesmundo}, \bibinfo{person}{Mona Attariyan}, {and} \bibinfo{person}{Sylvain
  Gelly}.} \bibinfo{year}{2019}\natexlab{}.
\newblock \showarticletitle{Parameter-efficient transfer learning for NLP}. In
  \bibinfo{booktitle}{\emph{International conference on machine learning}}.
  PMLR, \bibinfo{pages}{2790--2799}.
\newblock


\bibitem[Hu and Li(2024)]%
        {voltage}
\bibfield{author}{\bibinfo{person}{Chenghao Hu} {and} \bibinfo{person}{Baochun
  Li}.} \bibinfo{year}{2024}\natexlab{}.
\newblock \showarticletitle{When the Edge Meets Transformers: Distributed
  Inference with Transformer Models}. In \bibinfo{booktitle}{\emph{2024 IEEE
  44th International Conference on Distributed Computing Systems (ICDCS)}}.
  IEEE, \bibinfo{pages}{82--92}.
\newblock


\bibitem[Hu et~al\mbox{.}(2021)]%
        {lora}
\bibfield{author}{\bibinfo{person}{Edward~J Hu}, \bibinfo{person}{Yelong Shen},
  \bibinfo{person}{Phillip Wallis}, \bibinfo{person}{Zeyuan Allen-Zhu},
  \bibinfo{person}{Yuanzhi Li}, \bibinfo{person}{Shean Wang},
  \bibinfo{person}{Lu Wang}, {and} \bibinfo{person}{Weizhu Chen}.}
  \bibinfo{year}{2021}\natexlab{}.
\newblock \showarticletitle{Lora: Low-rank adaptation of large language
  models}.
\newblock \bibinfo{journal}{\emph{arXiv preprint arXiv:2106.09685}}
  (\bibinfo{year}{2021}).
\newblock


\bibitem[Jiang et~al\mbox{.}(2024)]%
        {mixtral}
\bibfield{author}{\bibinfo{person}{Albert~Q. Jiang}, \bibinfo{person}{Alexandre
  Sablayrolles}, \bibinfo{person}{Antoine Roux}, \bibinfo{person}{Arthur
  Mensch}, \bibinfo{person}{Blanche Savary}, \bibinfo{person}{Chris Bamford},
  \bibinfo{person}{Devendra~Singh Chaplot}, \bibinfo{person}{Diego de~las
  Casas}, \bibinfo{person}{Emma~Bou Hanna}, \bibinfo{person}{Florian Bressand},
  \bibinfo{person}{Gianna Lengyel}, \bibinfo{person}{Guillaume Bour},
  \bibinfo{person}{Guillaume Lample}, \bibinfo{person}{Lélio~Renard Lavaud},
  \bibinfo{person}{Lucile Saulnier}, \bibinfo{person}{Marie-Anne Lachaux},
  \bibinfo{person}{Pierre Stock}, \bibinfo{person}{Sandeep Subramanian},
  \bibinfo{person}{Sophia Yang}, \bibinfo{person}{Szymon Antoniak},
  \bibinfo{person}{Teven~Le Scao}, \bibinfo{person}{Théophile Gervet},
  \bibinfo{person}{Thibaut Lavril}, \bibinfo{person}{Thomas Wang},
  \bibinfo{person}{Timothée Lacroix}, {and} \bibinfo{person}{William~El
  Sayed}.} \bibinfo{year}{2024}\natexlab{}.
\newblock \bibinfo{title}{Mixtral of Experts}.
\newblock
\showeprint[arxiv]{2401.04088}~[cs.LG]
\urldef\tempurl%
\url{https://arxiv.org/abs/2401.04088}
\showURL{%
\tempurl}


\bibitem[Jiang et~al\mbox{.}(2023)]%
        {lion}
\bibfield{author}{\bibinfo{person}{Yuxin Jiang}, \bibinfo{person}{Chunkit
  Chan}, \bibinfo{person}{Mingyang Chen}, {and} \bibinfo{person}{Wei Wang}.}
  \bibinfo{year}{2023}\natexlab{}.
\newblock \showarticletitle{Lion: Adversarial distillation of proprietary large
  language models}.
\newblock \bibinfo{journal}{\emph{arXiv preprint arXiv:2305.12870}}
  (\bibinfo{year}{2023}).
\newblock


\bibitem[Josuttis(2012)]%
        {stl}
\bibfield{author}{\bibinfo{person}{Nicolai~M Josuttis}.}
  \bibinfo{year}{2012}\natexlab{}.
\newblock \showarticletitle{The C++ standard library: a tutorial and
  reference}.
\newblock  (\bibinfo{year}{2012}).
\newblock


\bibitem[Kwiatkowski et~al\mbox{.}(2019)]%
        {benchmark-qa}
\bibfield{author}{\bibinfo{person}{Tom Kwiatkowski},
  \bibinfo{person}{Jennimaria Palomaki}, \bibinfo{person}{Olivia Redfield},
  \bibinfo{person}{Michael Collins}, \bibinfo{person}{Ankur Parikh},
  \bibinfo{person}{Chris Alberti}, \bibinfo{person}{Danielle Epstein},
  \bibinfo{person}{Illia Polosukhin}, \bibinfo{person}{Jacob Devlin},
  \bibinfo{person}{Kenton Lee}, {et~al\mbox{.}}}
  \bibinfo{year}{2019}\natexlab{}.
\newblock \showarticletitle{Natural questions: a benchmark for question
  answering research}.
\newblock \bibinfo{journal}{\emph{Transactions of the Association for
  Computational Linguistics}}  \bibinfo{volume}{7} (\bibinfo{year}{2019}),
  \bibinfo{pages}{453--466}.
\newblock


\bibitem[Kwon et~al\mbox{.}(2023)]%
        {vllm}
\bibfield{author}{\bibinfo{person}{Woosuk Kwon}, \bibinfo{person}{Zhuohan Li},
  \bibinfo{person}{Siyuan Zhuang}, \bibinfo{person}{Ying Sheng},
  \bibinfo{person}{Lianmin Zheng}, \bibinfo{person}{Cody~Hao Yu},
  \bibinfo{person}{Joseph Gonzalez}, \bibinfo{person}{Hao Zhang}, {and}
  \bibinfo{person}{Ion Stoica}.} \bibinfo{year}{2023}\natexlab{}.
\newblock \showarticletitle{Efficient memory management for large language
  model serving with pagedattention}. In \bibinfo{booktitle}{\emph{Proceedings
  of the 29th Symposium on Operating Systems Principles}}.
  \bibinfo{pages}{611--626}.
\newblock


\bibitem[Lee et~al\mbox{.}(2024)]%
        {infinigen}
\bibfield{author}{\bibinfo{person}{Wonbeom Lee}, \bibinfo{person}{Jungi Lee},
  \bibinfo{person}{Junghwan Seo}, {and} \bibinfo{person}{Jaewoong Sim}.}
  \bibinfo{year}{2024}\natexlab{}.
\newblock \showarticletitle{$\{$InfiniGen$\}$: Efficient generative inference
  of large language models with dynamic $\{$KV$\}$ cache management}. In
  \bibinfo{booktitle}{\emph{18th USENIX Symposium on Operating Systems Design
  and Implementation (OSDI 24)}}. \bibinfo{pages}{155--172}.
\newblock


\bibitem[Lester et~al\mbox{.}(2021a)]%
        {prompt-peft}
\bibfield{author}{\bibinfo{person}{Brian Lester}, \bibinfo{person}{Rami
  Al-Rfou}, {and} \bibinfo{person}{Noah Constant}.}
  \bibinfo{year}{2021}\natexlab{a}.
\newblock \showarticletitle{The power of scale for parameter-efficient prompt
  tuning}.
\newblock \bibinfo{journal}{\emph{arXiv preprint arXiv:2104.08691}}
  (\bibinfo{year}{2021}).
\newblock


\bibitem[Lester et~al\mbox{.}(2021b)]%
        {prompt-tuning}
\bibfield{author}{\bibinfo{person}{Brian Lester}, \bibinfo{person}{Rami
  Al-Rfou}, {and} \bibinfo{person}{Noah Constant}.}
  \bibinfo{year}{2021}\natexlab{b}.
\newblock \showarticletitle{The power of scale for parameter-efficient prompt
  tuning}.
\newblock \bibinfo{journal}{\emph{arXiv preprint arXiv:2104.08691}}
  (\bibinfo{year}{2021}).
\newblock


\bibitem[Li et~al\mbox{.}(2024a)]%
        {cara-serve}
\bibfield{author}{\bibinfo{person}{Suyi Li}, \bibinfo{person}{Hanfeng Lu},
  \bibinfo{person}{Tianyuan Wu}, \bibinfo{person}{Minchen Yu},
  \bibinfo{person}{Qizhen Weng}, \bibinfo{person}{Xusheng Chen},
  \bibinfo{person}{Yizhou Shan}, \bibinfo{person}{Binhang Yuan}, {and}
  \bibinfo{person}{Wei Wang}.} \bibinfo{year}{2024}\natexlab{a}.
\newblock \bibinfo{title}{CaraServe: CPU-Assisted and Rank-Aware LoRA Serving
  for Generative LLM Inference}.
\newblock
\showeprint[arxiv]{2401.11240}~[cs.DC]
\urldef\tempurl%
\url{https://arxiv.org/abs/2401.11240}
\showURL{%
\tempurl}


\bibitem[Li and Liang(2021)]%
        {prefix-tuning}
\bibfield{author}{\bibinfo{person}{Xiang~Lisa Li} {and} \bibinfo{person}{Percy
  Liang}.} \bibinfo{year}{2021}\natexlab{}.
\newblock \showarticletitle{Prefix-tuning: Optimizing continuous prompts for
  generation}.
\newblock \bibinfo{journal}{\emph{arXiv preprint arXiv:2101.00190}}
  (\bibinfo{year}{2021}).
\newblock


\bibitem[Li et~al\mbox{.}(2023a)]%
        {chatdoctor}
\bibfield{author}{\bibinfo{person}{Yunxiang Li}, \bibinfo{person}{Zihan Li},
  \bibinfo{person}{Kai Zhang}, \bibinfo{person}{Ruilong Dan},
  \bibinfo{person}{Steve Jiang}, {and} \bibinfo{person}{You Zhang}.}
  \bibinfo{year}{2023}\natexlab{a}.
\newblock \showarticletitle{Chatdoctor: A medical chat model fine-tuned on a
  large language model meta-ai (llama) using medical domain knowledge}.
\newblock \bibinfo{journal}{\emph{Cureus}} \bibinfo{volume}{15},
  \bibinfo{number}{6} (\bibinfo{year}{2023}).
\newblock


\bibitem[Li et~al\mbox{.}(2024b)]%
        {personal-agent}
\bibfield{author}{\bibinfo{person}{Yuanchun Li}, \bibinfo{person}{Hao Wen},
  \bibinfo{person}{Weijun Wang}, \bibinfo{person}{Xiangyu Li},
  \bibinfo{person}{Yizhen Yuan}, \bibinfo{person}{Guohong Liu},
  \bibinfo{person}{Jiacheng Liu}, \bibinfo{person}{Wenxing Xu},
  \bibinfo{person}{Xiang Wang}, \bibinfo{person}{Yi Sun}, {et~al\mbox{.}}}
  \bibinfo{year}{2024}\natexlab{b}.
\newblock \showarticletitle{Personal llm agents: Insights and survey about the
  capability, efficiency and security}.
\newblock \bibinfo{journal}{\emph{arXiv preprint arXiv:2401.05459}}
  (\bibinfo{year}{2024}).
\newblock


\bibitem[Li et~al\mbox{.}(2023b)]%
        {alpaserve}
\bibfield{author}{\bibinfo{person}{Zhuohan Li}, \bibinfo{person}{Lianmin
  Zheng}, \bibinfo{person}{Yinmin Zhong}, \bibinfo{person}{Vincent Liu},
  \bibinfo{person}{Ying Sheng}, \bibinfo{person}{Xin Jin},
  \bibinfo{person}{Yanping Huang}, \bibinfo{person}{Zhifeng Chen},
  \bibinfo{person}{Hao Zhang}, \bibinfo{person}{Joseph~E Gonzalez},
  {et~al\mbox{.}}} \bibinfo{year}{2023}\natexlab{b}.
\newblock \showarticletitle{$\{$AlpaServe$\}$: Statistical multiplexing with
  model parallelism for deep learning serving}. In
  \bibinfo{booktitle}{\emph{17th USENIX Symposium on Operating Systems Design
  and Implementation (OSDI 23)}}. \bibinfo{pages}{663--679}.
\newblock


\bibitem[Lian et~al\mbox{.}(2023)]%
        {llm-video}
\bibfield{author}{\bibinfo{person}{Long Lian}, \bibinfo{person}{Baifeng Shi},
  \bibinfo{person}{Adam Yala}, \bibinfo{person}{Trevor Darrell}, {and}
  \bibinfo{person}{Boyi Li}.} \bibinfo{year}{2023}\natexlab{}.
\newblock \showarticletitle{Llm-grounded video diffusion models}.
\newblock \bibinfo{journal}{\emph{arXiv preprint arXiv:2309.17444}}
  (\bibinfo{year}{2023}).
\newblock


\bibitem[Lin et~al\mbox{.}(2024)]%
        {awq}
\bibfield{author}{\bibinfo{person}{Ji Lin}, \bibinfo{person}{Jiaming Tang},
  \bibinfo{person}{Haotian Tang}, \bibinfo{person}{Shang Yang},
  \bibinfo{person}{Wei-Ming Chen}, \bibinfo{person}{Wei-Chen Wang},
  \bibinfo{person}{Guangxuan Xiao}, \bibinfo{person}{Xingyu Dang},
  \bibinfo{person}{Chuang Gan}, {and} \bibinfo{person}{Song Han}.}
  \bibinfo{year}{2024}\natexlab{}.
\newblock \showarticletitle{Awq: Activation-aware weight quantization for
  on-device llm compression and acceleration}.
\newblock \bibinfo{journal}{\emph{Proceedings of Machine Learning and Systems}}
   \bibinfo{volume}{6} (\bibinfo{year}{2024}), \bibinfo{pages}{87--100}.
\newblock


\bibitem[Liu and Low(2023)]%
        {goat}
\bibfield{author}{\bibinfo{person}{Tiedong Liu} {and} \bibinfo{person}{Bryan
  Kian~Hsiang Low}.} \bibinfo{year}{2023}\natexlab{}.
\newblock \showarticletitle{Goat: Fine-tuned llama outperforms gpt-4 on
  arithmetic tasks}.
\newblock \bibinfo{journal}{\emph{arXiv preprint arXiv:2305.14201}}
  (\bibinfo{year}{2023}).
\newblock


\bibitem[Liu et~al\mbox{.}(2021a)]%
        {p-tuning}
\bibfield{author}{\bibinfo{person}{Xiao Liu}, \bibinfo{person}{Kaixuan Ji},
  \bibinfo{person}{Yicheng Fu}, \bibinfo{person}{Weng~Lam Tam},
  \bibinfo{person}{Zhengxiao Du}, \bibinfo{person}{Zhilin Yang}, {and}
  \bibinfo{person}{Jie Tang}.} \bibinfo{year}{2021}\natexlab{a}.
\newblock \showarticletitle{P-tuning v2: Prompt tuning can be comparable to
  fine-tuning universally across scales and tasks}.
\newblock \bibinfo{journal}{\emph{arXiv preprint arXiv:2110.07602}}
  (\bibinfo{year}{2021}).
\newblock


\bibitem[Liu et~al\mbox{.}(2021b)]%
        {swin}
\bibfield{author}{\bibinfo{person}{Ze Liu}, \bibinfo{person}{Yutong Lin},
  \bibinfo{person}{Yue Cao}, \bibinfo{person}{Han Hu}, \bibinfo{person}{Yixuan
  Wei}, \bibinfo{person}{Zheng Zhang}, \bibinfo{person}{Stephen Lin}, {and}
  \bibinfo{person}{Baining Guo}.} \bibinfo{year}{2021}\natexlab{b}.
\newblock \showarticletitle{Swin transformer: Hierarchical vision transformer
  using shifted windows}. In \bibinfo{booktitle}{\emph{Proceedings of the
  IEEE/CVF international conference on computer vision}}.
  \bibinfo{pages}{10012--10022}.
\newblock


\bibitem[Luo et~al\mbox{.}(2022)]%
        {chatbot-luo}
\bibfield{author}{\bibinfo{person}{Bei Luo}, \bibinfo{person}{Raymond~YK Lau},
  \bibinfo{person}{Chunping Li}, {and} \bibinfo{person}{Yain-Whar Si}.}
  \bibinfo{year}{2022}\natexlab{}.
\newblock \showarticletitle{A critical review of state-of-the-art chatbot
  designs and applications}.
\newblock \bibinfo{journal}{\emph{Wiley Interdisciplinary Reviews: Data Mining
  and Knowledge Discovery}} \bibinfo{volume}{12}, \bibinfo{number}{1}
  (\bibinfo{year}{2022}), \bibinfo{pages}{e1434}.
\newblock


\bibitem[Ma et~al\mbox{.}(2023)]%
        {llm-pruner}
\bibfield{author}{\bibinfo{person}{Xinyin Ma}, \bibinfo{person}{Gongfan Fang},
  {and} \bibinfo{person}{Xinchao Wang}.} \bibinfo{year}{2023}\natexlab{}.
\newblock \showarticletitle{Llm-pruner: On the structural pruning of large
  language models}.
\newblock \bibinfo{journal}{\emph{Advances in neural information processing
  systems}}  \bibinfo{volume}{36} (\bibinfo{year}{2023}),
  \bibinfo{pages}{21702--21720}.
\newblock


\bibitem[Mangrulkar et~al\mbox{.}(2022)]%
        {peft-library}
\bibfield{author}{\bibinfo{person}{Sourab Mangrulkar}, \bibinfo{person}{Sylvain
  Gugger}, \bibinfo{person}{Lysandre Debut}, \bibinfo{person}{Younes Belkada},
  \bibinfo{person}{Sayak Paul}, {and} \bibinfo{person}{Benjamin Bossan}.}
  \bibinfo{year}{2022}\natexlab{}.
\newblock \bibinfo{title}{PEFT: State-of-the-art Parameter-Efficient
  Fine-Tuning methods}.
\newblock \bibinfo{howpublished}{\url{https://github.com/huggingface/peft}}.
\newblock


\bibitem[Mehta et~al\mbox{.}(2024)]%
        {openelm}
\bibfield{author}{\bibinfo{person}{Sachin Mehta},
  \bibinfo{person}{Mohammad~Hossein Sekhavat}, \bibinfo{person}{Qingqing Cao},
  \bibinfo{person}{Maxwell Horton}, \bibinfo{person}{Yanzi Jin},
  \bibinfo{person}{Chenfan Sun}, \bibinfo{person}{Iman Mirzadeh},
  \bibinfo{person}{Mahyar Najibi}, \bibinfo{person}{Dmitry Belenko},
  \bibinfo{person}{Peter Zatloukal}, {and} \bibinfo{person}{Mohammad
  Rastegari}.} \bibinfo{year}{2024}\natexlab{}.
\newblock \bibinfo{title}{OpenELM: An Efficient Language Model Family with Open
  Training and Inference Framework}.
\newblock
\showeprint[arxiv]{2404.14619}~[cs.CL]
\urldef\tempurl%
\url{https://arxiv.org/abs/2404.14619}
\showURL{%
\tempurl}


\bibitem[Mi et~al\mbox{.}(2024)]%
        {vlora}
\bibfield{author}{\bibinfo{person}{Liang Mi}, \bibinfo{person}{Weijun Wang},
  \bibinfo{person}{Wenming Tu}, \bibinfo{person}{Qingfeng He},
  \bibinfo{person}{Rui Kong}, \bibinfo{person}{Xinyu Fang},
  \bibinfo{person}{Yazhu Dong}, \bibinfo{person}{Yikang Zhang},
  \bibinfo{person}{Yunchun Li}, \bibinfo{person}{Meng Li}, {et~al\mbox{.}}}
  \bibinfo{year}{2024}\natexlab{}.
\newblock \showarticletitle{V-LoRA: An Efficient and Flexible System Boosts
  Vision Applications with LoRA LMM}.
\newblock \bibinfo{journal}{\emph{arXiv preprint arXiv:2411.00915}}
  (\bibinfo{year}{2024}).
\newblock


\bibitem[Microsoft(2023)]%
        {deepspeedmii}
\bibfield{author}{\bibinfo{person}{Microsoft}.}
  \bibinfo{year}{2023}\natexlab{}.
\newblock \bibinfo{title}{DeepSpeed-MII: DeepSpeed Model Implementations for
  Inference}.
\newblock
  \bibinfo{howpublished}{\url{https://github.com/microsoft/DeepSpeed-MII}}.
\newblock
\newblock
\shownote{Accessed: [Insert date of access]}.


\bibitem[{MLC team}(2025)]%
        {mlc-llm}
\bibfield{author}{\bibinfo{person}{{MLC team}}.}
  \bibinfo{year}{2023-2025}\natexlab{}.
\newblock \bibinfo{booktitle}{\emph{{MLC-LLM}}}.
\newblock
\urldef\tempurl%
\url{https://github.com/mlc-ai/mlc-llm}
\showURL{%
\tempurl}


\bibitem[Mysore et~al\mbox{.}(2023)]%
        {writing-assistant}
\bibfield{author}{\bibinfo{person}{Sheshera Mysore}, \bibinfo{person}{Zhuoran
  Lu}, \bibinfo{person}{Mengting Wan}, \bibinfo{person}{Longqi Yang},
  \bibinfo{person}{Steve Menezes}, \bibinfo{person}{Tina Baghaee},
  \bibinfo{person}{Emmanuel~Barajas Gonzalez}, \bibinfo{person}{Jennifer
  Neville}, {and} \bibinfo{person}{Tara Safavi}.}
  \bibinfo{year}{2023}\natexlab{}.
\newblock \showarticletitle{Pearl: Personalizing large language model writing
  assistants with generation-calibrated retrievers}.
\newblock \bibinfo{journal}{\emph{arXiv preprint arXiv:2311.09180}}
  (\bibinfo{year}{2023}).
\newblock


\bibitem[Nallapati et~al\mbox{.}(2016)]%
        {abstractive-summary}
\bibfield{author}{\bibinfo{person}{Ramesh Nallapati}, \bibinfo{person}{Bowen
  Zhou}, \bibinfo{person}{Caglar Gulcehre}, \bibinfo{person}{Bing Xiang},
  {et~al\mbox{.}}} \bibinfo{year}{2016}\natexlab{}.
\newblock \showarticletitle{Abstractive text summarization using
  sequence-to-sequence rnns and beyond}.
\newblock \bibinfo{journal}{\emph{arXiv preprint arXiv:1602.06023}}
  (\bibinfo{year}{2016}).
\newblock


\bibitem[OpenAI(2024)]%
        {gpt4}
\bibfield{author}{\bibinfo{person}{OpenAI}.} \bibinfo{year}{2024}\natexlab{}.
\newblock \bibinfo{title}{GPT-4 Technical Report}.
\newblock
\showeprint[arxiv]{2303.08774}~[cs.CL]
\urldef\tempurl%
\url{https://arxiv.org/abs/2303.08774}
\showURL{%
\tempurl}


\bibitem[Ouyang et~al\mbox{.}(2024)]%
        {chatgpt-code}
\bibfield{author}{\bibinfo{person}{Shuyin Ouyang}, \bibinfo{person}{Jie~M.
  Zhang}, \bibinfo{person}{Mark Harman}, {and} \bibinfo{person}{Meng Wang}.}
  \bibinfo{year}{2024}\natexlab{}.
\newblock \showarticletitle{An Empirical Study of the Non-determinism of
  ChatGPT in Code Generation}.
\newblock \bibinfo{journal}{\emph{ACM Transactions on Software Engineering and
  Methodology}} (\bibinfo{date}{Sept.} \bibinfo{year}{2024}).
\newblock
\showISSN{1557-7392}
\urldef\tempurl%
\url{https://doi.org/10.1145/3697010}
\showDOI{\tempurl}


\bibitem[Paolieri et~al\mbox{.}(2011)]%
        {ia3}
\bibfield{author}{\bibinfo{person}{Marco Paolieri}, \bibinfo{person}{Eduardo
  Qui{\~n}ones}, \bibinfo{person}{Francisco~J Cazorla},
  \bibinfo{person}{Robert~I Davis}, {and} \bibinfo{person}{Mateo Valero}.}
  \bibinfo{year}{2011}\natexlab{}.
\newblock \showarticletitle{IA\^{} 3: An interference aware allocation
  algorithm for multicore hard real-time systems}. In
  \bibinfo{booktitle}{\emph{2011 17th IEEE Real-Time and Embedded Technology
  and Applications Symposium}}. IEEE, \bibinfo{pages}{280--290}.
\newblock


\bibitem[Paszke et~al\mbox{.}(2017)]%
        {pytorch}
\bibfield{author}{\bibinfo{person}{Adam Paszke}, \bibinfo{person}{Sam Gross},
  \bibinfo{person}{Soumith Chintala}, \bibinfo{person}{Gregory Chanan},
  \bibinfo{person}{Edward Yang}, \bibinfo{person}{Zachary DeVito},
  \bibinfo{person}{Zeming Lin}, \bibinfo{person}{Alban Desmaison},
  \bibinfo{person}{Luca Antiga}, {and} \bibinfo{person}{Adam Lerer}.}
  \bibinfo{year}{2017}\natexlab{}.
\newblock \showarticletitle{Automatic differentiation in PyTorch}.
\newblock  (\bibinfo{year}{2017}).
\newblock


\bibitem[Peterson and Silberschatz(1985)]%
        {lru}
\bibfield{author}{\bibinfo{person}{James~L Peterson} {and}
  \bibinfo{person}{Abraham Silberschatz}.} \bibinfo{year}{1985}\natexlab{}.
\newblock \bibinfo{booktitle}{\emph{Operating system concepts}}.
\newblock \bibinfo{publisher}{Addison-Wesley Longman Publishing Co., Inc.}
\newblock


\bibitem[Rein et~al\mbox{.}(2023)]%
        {gpqa}
\bibfield{author}{\bibinfo{person}{David Rein}, \bibinfo{person}{Betty~Li Hou},
  \bibinfo{person}{Asa~Cooper Stickland}, \bibinfo{person}{Jackson Petty},
  \bibinfo{person}{Richard~Yuanzhe Pang}, \bibinfo{person}{Julien Dirani},
  \bibinfo{person}{Julian Michael}, {and} \bibinfo{person}{Samuel~R. Bowman}.}
  \bibinfo{year}{2023}\natexlab{}.
\newblock \bibinfo{title}{GPQA: A Graduate-Level Google-Proof Q\&A Benchmark}.
\newblock
\showeprint[arxiv]{2311.12022}~[cs.AI]
\urldef\tempurl%
\url{https://arxiv.org/abs/2311.12022}
\showURL{%
\tempurl}


\bibitem[Roziere et~al\mbox{.}(2023)]%
        {code-llama}
\bibfield{author}{\bibinfo{person}{Baptiste Roziere}, \bibinfo{person}{Jonas
  Gehring}, \bibinfo{person}{Fabian Gloeckle}, \bibinfo{person}{Sten Sootla},
  \bibinfo{person}{Itai Gat}, \bibinfo{person}{Xiaoqing~Ellen Tan},
  \bibinfo{person}{Yossi Adi}, \bibinfo{person}{Jingyu Liu},
  \bibinfo{person}{Romain Sauvestre}, \bibinfo{person}{Tal Remez},
  {et~al\mbox{.}}} \bibinfo{year}{2023}\natexlab{}.
\newblock \showarticletitle{Code llama: Open foundation models for code}.
\newblock \bibinfo{journal}{\emph{arXiv preprint arXiv:2308.12950}}
  (\bibinfo{year}{2023}).
\newblock


\bibitem[Schick and Sch{\"u}tze(2020)]%
        {text-classification}
\bibfield{author}{\bibinfo{person}{Timo Schick} {and} \bibinfo{person}{Hinrich
  Sch{\"u}tze}.} \bibinfo{year}{2020}\natexlab{}.
\newblock \showarticletitle{Exploiting cloze questions for few shot text
  classification and natural language inference}.
\newblock \bibinfo{journal}{\emph{arXiv preprint arXiv:2001.07676}}
  (\bibinfo{year}{2020}).
\newblock


\bibitem[Shen et~al\mbox{.}(2022)]%
        {soter}
\bibfield{author}{\bibinfo{person}{Tianxiang Shen}, \bibinfo{person}{Ji Qi},
  \bibinfo{person}{Jianyu Jiang}, \bibinfo{person}{Xian Wang},
  \bibinfo{person}{Siyuan Wen}, \bibinfo{person}{Xusheng Chen},
  \bibinfo{person}{Shixiong Zhao}, \bibinfo{person}{Sen Wang},
  \bibinfo{person}{Li Chen}, \bibinfo{person}{Xiapu Luo},
  \bibinfo{person}{Fengwei Zhang}, {and} \bibinfo{person}{Heming Cui}.}
  \bibinfo{year}{2022}\natexlab{}.
\newblock \showarticletitle{{SOTER}: Guarding Black-box Inference for General
  Neural Networks at the Edge}. In \bibinfo{booktitle}{\emph{2022 USENIX Annual
  Technical Conference (USENIX ATC 22)}}. \bibinfo{publisher}{USENIX
  Association}, \bibinfo{address}{Carlsbad, CA}, \bibinfo{pages}{723--738}.
\newblock
\showISBNx{978-1-939133-29-68}
\urldef\tempurl%
\url{https://www.usenix.org/conference/atc22/presentation/shen}
\showURL{%
\tempurl}


\bibitem[Sheng et~al\mbox{.}(2023)]%
        {slora}
\bibfield{author}{\bibinfo{person}{Ying Sheng}, \bibinfo{person}{Shiyi Cao},
  \bibinfo{person}{Dacheng Li}, \bibinfo{person}{Coleman Hooper},
  \bibinfo{person}{Nicholas Lee}, \bibinfo{person}{Shuo Yang},
  \bibinfo{person}{Christopher Chou}, \bibinfo{person}{Banghua Zhu},
  \bibinfo{person}{Lianmin Zheng}, \bibinfo{person}{Kurt Keutzer},
  {et~al\mbox{.}}} \bibinfo{year}{2023}\natexlab{}.
\newblock \showarticletitle{S-lora: Serving thousands of concurrent lora
  adapters}.
\newblock \bibinfo{journal}{\emph{arXiv preprint arXiv:2311.03285}}
  (\bibinfo{year}{2023}).
\newblock


\bibitem[Shu et~al\mbox{.}(2023)]%
        {llasm}
\bibfield{author}{\bibinfo{person}{Yu Shu}, \bibinfo{person}{Siwei Dong},
  \bibinfo{person}{Guangyao Chen}, \bibinfo{person}{Wenhao Huang},
  \bibinfo{person}{Ruihua Zhang}, \bibinfo{person}{Daochen Shi},
  \bibinfo{person}{Qiqi Xiang}, {and} \bibinfo{person}{Yemin Shi}.}
  \bibinfo{year}{2023}\natexlab{}.
\newblock \showarticletitle{Llasm: Large language and speech model}.
\newblock \bibinfo{journal}{\emph{arXiv preprint arXiv:2308.15930}}
  (\bibinfo{year}{2023}).
\newblock


\bibitem[Song et~al\mbox{.}(2024)]%
        {powerinfer}
\bibfield{author}{\bibinfo{person}{Yixin Song}, \bibinfo{person}{Zeyu Mi},
  \bibinfo{person}{Haotong Xie}, {and} \bibinfo{person}{Haibo Chen}.}
  \bibinfo{year}{2024}\natexlab{}.
\newblock \showarticletitle{Powerinfer: Fast large language model serving with
  a consumer-grade gpu}. In \bibinfo{booktitle}{\emph{Proceedings of the ACM
  SIGOPS 30th Symposium on Operating Systems Principles}}.
  \bibinfo{pages}{590--606}.
\newblock


\bibitem[Stahlberg(2020)]%
        {translation}
\bibfield{author}{\bibinfo{person}{Felix Stahlberg}.}
  \bibinfo{year}{2020}\natexlab{}.
\newblock \showarticletitle{Neural machine translation: A review}.
\newblock \bibinfo{journal}{\emph{Journal of Artificial Intelligence Research}}
   \bibinfo{volume}{69} (\bibinfo{year}{2020}), \bibinfo{pages}{343--418}.
\newblock


\bibitem[Sun et~al\mbox{.}(2024)]%
        {llumnix}
\bibfield{author}{\bibinfo{person}{Biao Sun}, \bibinfo{person}{Ziming Huang},
  \bibinfo{person}{Hanyu Zhao}, \bibinfo{person}{Wencong Xiao},
  \bibinfo{person}{Xinyi Zhang}, \bibinfo{person}{Yong Li}, {and}
  \bibinfo{person}{Wei Lin}.} \bibinfo{year}{2024}\natexlab{}.
\newblock \showarticletitle{Llumnix: Dynamic Scheduling for Large Language
  Model Serving}.
\newblock \bibinfo{journal}{\emph{arXiv preprint arXiv:2406.03243}}
  (\bibinfo{year}{2024}).
\newblock


\bibitem[Suzgun et~al\mbox{.}(2022)]%
        {bbh}
\bibfield{author}{\bibinfo{person}{Mirac Suzgun}, \bibinfo{person}{Nathan
  Scales}, \bibinfo{person}{Nathanael Schärli}, \bibinfo{person}{Sebastian
  Gehrmann}, \bibinfo{person}{Yi Tay}, \bibinfo{person}{Hyung~Won Chung},
  \bibinfo{person}{Aakanksha Chowdhery}, \bibinfo{person}{Quoc~V. Le},
  \bibinfo{person}{Ed~H. Chi}, \bibinfo{person}{Denny Zhou}, {and}
  \bibinfo{person}{Jason Wei}.} \bibinfo{year}{2022}\natexlab{}.
\newblock \bibinfo{title}{Challenging BIG-Bench Tasks and Whether
  Chain-of-Thought Can Solve Them}.
\newblock
\showeprint[arxiv]{2210.09261}~[cs.CL]
\urldef\tempurl%
\url{https://arxiv.org/abs/2210.09261}
\showURL{%
\tempurl}


\bibitem[Tang et~al\mbox{.}(2024)]%
        {mathscale}
\bibfield{author}{\bibinfo{person}{Zhengyang Tang}, \bibinfo{person}{Xingxing
  Zhang}, \bibinfo{person}{Benyou Wang}, {and} \bibinfo{person}{Furu Wei}.}
  \bibinfo{year}{2024}\natexlab{}.
\newblock \bibinfo{title}{MathScale: Scaling Instruction Tuning for
  Mathematical Reasoning}.
\newblock
\showeprint[arxiv]{2403.02884}~[cs.CL]
\urldef\tempurl%
\url{https://arxiv.org/abs/2403.02884}
\showURL{%
\tempurl}


\bibitem[Taori et~al\mbox{.}(2023)]%
        {alpaca}
\bibfield{author}{\bibinfo{person}{Rohan Taori}, \bibinfo{person}{Ishaan
  Gulrajani}, \bibinfo{person}{Tianyi Zhang}, \bibinfo{person}{Yann Dubois},
  \bibinfo{person}{Xuechen Li}, \bibinfo{person}{Carlos Guestrin},
  \bibinfo{person}{Percy Liang}, {and} \bibinfo{person}{Tatsunori~B
  Hashimoto}.} \bibinfo{year}{2023}\natexlab{}.
\newblock \showarticletitle{Alpaca: A strong, replicable instruction-following
  model}.
\newblock \bibinfo{journal}{\emph{Stanford Center for Research on Foundation
  Models. https://crfm. stanford. edu/2023/03/13/alpaca. html}}
  \bibinfo{volume}{3}, \bibinfo{number}{6} (\bibinfo{year}{2023}),
  \bibinfo{pages}{7}.
\newblock


\bibitem[Team(2024)]%
        {gemini}
\bibfield{author}{\bibinfo{person}{Gemini Team}.}
  \bibinfo{year}{2024}\natexlab{}.
\newblock \bibinfo{title}{Gemini: A Family of Highly Capable Multimodal
  Models}.
\newblock
\showeprint[arxiv]{2312.11805}~[cs.CL]
\urldef\tempurl%
\url{https://arxiv.org/abs/2312.11805}
\showURL{%
\tempurl}


\bibitem[Toshniwal et~al\mbox{.}(2024)]%
        {openmath2}
\bibfield{author}{\bibinfo{person}{Shubham Toshniwal}, \bibinfo{person}{Wei
  Du}, \bibinfo{person}{Ivan Moshkov}, \bibinfo{person}{Branislav Kisacanin},
  \bibinfo{person}{Alexan Ayrapetyan}, {and} \bibinfo{person}{Igor Gitman}.}
  \bibinfo{year}{2024}\natexlab{}.
\newblock \showarticletitle{OpenMathInstruct-2: Accelerating AI for Math with
  Massive Open-Source Instruction Data}.
\newblock \bibinfo{journal}{\emph{arXiv preprint arXiv:2410.01560}}
  (\bibinfo{year}{2024}).
\newblock


\bibitem[Touvron et~al\mbox{.}(2023)]%
        {llama}
\bibfield{author}{\bibinfo{person}{Hugo Touvron}, \bibinfo{person}{Thibaut
  Lavril}, \bibinfo{person}{Gautier Izacard}, \bibinfo{person}{Xavier
  Martinet}, \bibinfo{person}{Marie-Anne Lachaux}, \bibinfo{person}{Timothée
  Lacroix}, \bibinfo{person}{Baptiste Rozière}, \bibinfo{person}{Naman Goyal},
  \bibinfo{person}{Eric Hambro}, \bibinfo{person}{Faisal Azhar},
  \bibinfo{person}{Aurelien Rodriguez}, \bibinfo{person}{Armand Joulin},
  \bibinfo{person}{Edouard Grave}, {and} \bibinfo{person}{Guillaume Lample}.}
  \bibinfo{year}{2023}\natexlab{}.
\newblock \bibinfo{title}{LLaMA: Open and Efficient Foundation Language
  Models}.
\newblock
\showeprint[arxiv]{2302.13971}~[cs.CL]
\urldef\tempurl%
\url{https://arxiv.org/abs/2302.13971}
\showURL{%
\tempurl}


\bibitem[Vaswani(2017)]%
        {transformer}
\bibfield{author}{\bibinfo{person}{A Vaswani}.}
  \bibinfo{year}{2017}\natexlab{}.
\newblock \showarticletitle{Attention is all you need}.
\newblock \bibinfo{journal}{\emph{Advances in Neural Information Processing
  Systems}} (\bibinfo{year}{2017}).
\newblock


\bibitem[Wang et~al\mbox{.}(2020)]%
        {minilm}
\bibfield{author}{\bibinfo{person}{Wenhui Wang}, \bibinfo{person}{Furu Wei},
  \bibinfo{person}{Li Dong}, \bibinfo{person}{Hangbo Bao}, \bibinfo{person}{Nan
  Yang}, {and} \bibinfo{person}{Ming Zhou}.} \bibinfo{year}{2020}\natexlab{}.
\newblock \showarticletitle{Minilm: Deep self-attention distillation for
  task-agnostic compression of pre-trained transformers}.
\newblock \bibinfo{journal}{\emph{Advances in neural information processing
  systems}}  \bibinfo{volume}{33} (\bibinfo{year}{2020}),
  \bibinfo{pages}{5776--5788}.
\newblock


\bibitem[Wang et~al\mbox{.}(2024a)]%
        {mmlu-pro}
\bibfield{author}{\bibinfo{person}{Yubo Wang}, \bibinfo{person}{Xueguang Ma},
  \bibinfo{person}{Ge Zhang}, \bibinfo{person}{Yuansheng Ni},
  \bibinfo{person}{Abhranil Chandra}, \bibinfo{person}{Shiguang Guo},
  \bibinfo{person}{Weiming Ren}, \bibinfo{person}{Aaran Arulraj},
  \bibinfo{person}{Xuan He}, \bibinfo{person}{Ziyan Jiang},
  \bibinfo{person}{Tianle Li}, \bibinfo{person}{Max Ku}, \bibinfo{person}{Kai
  Wang}, \bibinfo{person}{Alex Zhuang}, \bibinfo{person}{Rongqi Fan},
  \bibinfo{person}{Xiang Yue}, {and} \bibinfo{person}{Wenhu Chen}.}
  \bibinfo{year}{2024}\natexlab{a}.
\newblock \bibinfo{title}{MMLU-Pro: A More Robust and Challenging Multi-Task
  Language Understanding Benchmark}.
\newblock
\showeprint[arxiv]{2406.01574}~[cs.CL]
\urldef\tempurl%
\url{https://arxiv.org/abs/2406.01574}
\showURL{%
\tempurl}


\bibitem[Wang et~al\mbox{.}(2022)]%
        {clip-gen}
\bibfield{author}{\bibinfo{person}{Zihao Wang}, \bibinfo{person}{Wei Liu},
  \bibinfo{person}{Qian He}, \bibinfo{person}{Xinglong Wu}, {and}
  \bibinfo{person}{Zili Yi}.} \bibinfo{year}{2022}\natexlab{}.
\newblock \showarticletitle{Clip-gen: Language-free training of a text-to-image
  generator with clip}.
\newblock \bibinfo{journal}{\emph{arXiv preprint arXiv:2203.00386}}
  (\bibinfo{year}{2022}).
\newblock


\bibitem[Wang et~al\mbox{.}(2024b)]%
        {flora}
\bibfield{author}{\bibinfo{person}{Ziyao Wang}, \bibinfo{person}{Zheyu Shen},
  \bibinfo{person}{Yexiao He}, \bibinfo{person}{Guoheng Sun},
  \bibinfo{person}{Hongyi Wang}, \bibinfo{person}{Lingjuan Lyu}, {and}
  \bibinfo{person}{Ang Li}.} \bibinfo{year}{2024}\natexlab{b}.
\newblock \bibinfo{title}{FLoRA: Federated Fine-Tuning Large Language Models
  with Heterogeneous Low-Rank Adaptations}.
\newblock
\showeprint[arxiv]{2409.05976}~[cs.LG]
\urldef\tempurl%
\url{https://arxiv.org/abs/2409.05976}
\showURL{%
\tempurl}


\bibitem[Widyassari et~al\mbox{.}(2022)]%
        {summarization-review}
\bibfield{author}{\bibinfo{person}{Adhika~Pramita Widyassari},
  \bibinfo{person}{Supriadi Rustad}, \bibinfo{person}{Guruh~Fajar Shidik},
  \bibinfo{person}{Edi Noersasongko}, \bibinfo{person}{Abdul Syukur},
  \bibinfo{person}{Affandy Affandy}, {et~al\mbox{.}}}
  \bibinfo{year}{2022}\natexlab{}.
\newblock \showarticletitle{Review of automatic text summarization techniques
  \& methods}.
\newblock \bibinfo{journal}{\emph{Journal of King Saud University-Computer and
  Information Sciences}} \bibinfo{volume}{34}, \bibinfo{number}{4}
  (\bibinfo{year}{2022}), \bibinfo{pages}{1029--1046}.
\newblock


\bibitem[Wolf et~al\mbox{.}(2020)]%
        {transformers-library}
\bibfield{author}{\bibinfo{person}{Thomas Wolf}, \bibinfo{person}{Lysandre
  Debut}, \bibinfo{person}{Victor Sanh}, \bibinfo{person}{Julien Chaumond},
  \bibinfo{person}{Clement Delangue}, \bibinfo{person}{Anthony Moi},
  \bibinfo{person}{Pierric Cistac}, \bibinfo{person}{Tim Rault},
  \bibinfo{person}{Rémi Louf}, \bibinfo{person}{Morgan Funtowicz},
  \bibinfo{person}{Joe Davison}, \bibinfo{person}{Sam Shleifer},
  \bibinfo{person}{Patrick von Platen}, \bibinfo{person}{Clara Ma},
  \bibinfo{person}{Yacine Jernite}, \bibinfo{person}{Julien Plu},
  \bibinfo{person}{Canwen Xu}, \bibinfo{person}{Teven~Le Scao},
  \bibinfo{person}{Sylvain Gugger}, \bibinfo{person}{Mariama Drame},
  \bibinfo{person}{Quentin Lhoest}, {and} \bibinfo{person}{Alexander~M. Rush}.}
  \bibinfo{year}{2020}\natexlab{}.
\newblock \showarticletitle{Transformers: State-of-the-Art Natural Language
  Processing}. In \bibinfo{booktitle}{\emph{Proceedings of the 2020 Conference
  on Empirical Methods in Natural Language Processing: System Demonstrations}}.
  \bibinfo{publisher}{Association for Computational Linguistics},
  \bibinfo{address}{Online}, \bibinfo{pages}{38--45}.
\newblock
\urldef\tempurl%
\url{https://www.aclweb.org/anthology/2020.emnlp-demos.6}
\showURL{%
\tempurl}


\bibitem[Wu et~al\mbox{.}(2024)]%
        {dlora}
\bibfield{author}{\bibinfo{person}{Bingyang Wu}, \bibinfo{person}{Ruidong Zhu},
  \bibinfo{person}{Zili Zhang}, \bibinfo{person}{Peng Sun},
  \bibinfo{person}{Xuanzhe Liu}, {and} \bibinfo{person}{Xin Jin}.}
  \bibinfo{year}{2024}\natexlab{}.
\newblock \showarticletitle{$\{$dLoRA$\}$: Dynamically Orchestrating Requests
  and Adapters for $\{$LoRA$\}$$\{$LLM$\}$ Serving}. In
  \bibinfo{booktitle}{\emph{18th USENIX Symposium on Operating Systems Design
  and Implementation (OSDI 24)}}. \bibinfo{pages}{911--927}.
\newblock


\bibitem[Xia et~al\mbox{.}(2022)]%
        {cofi}
\bibfield{author}{\bibinfo{person}{Mengzhou Xia}, \bibinfo{person}{Zexuan
  Zhong}, {and} \bibinfo{person}{Danqi Chen}.} \bibinfo{year}{2022}\natexlab{}.
\newblock \showarticletitle{Structured pruning learns compact and accurate
  models}.
\newblock \bibinfo{journal}{\emph{arXiv preprint arXiv:2204.00408}}
  (\bibinfo{year}{2022}).
\newblock


\bibitem[Xiao et~al\mbox{.}(2023)]%
        {smoothquant}
\bibfield{author}{\bibinfo{person}{Guangxuan Xiao}, \bibinfo{person}{Ji Lin},
  \bibinfo{person}{Mickael Seznec}, \bibinfo{person}{Hao Wu},
  \bibinfo{person}{Julien Demouth}, {and} \bibinfo{person}{Song Han}.}
  \bibinfo{year}{2023}\natexlab{}.
\newblock \showarticletitle{Smoothquant: Accurate and efficient post-training
  quantization for large language models}. In
  \bibinfo{booktitle}{\emph{International Conference on Machine Learning}}.
  PMLR, \bibinfo{pages}{38087--38099}.
\newblock


\bibitem[Yang et~al\mbox{.}(2025)]%
        {lserve}
\bibfield{author}{\bibinfo{person}{Shang Yang}, \bibinfo{person}{Junxian Guo},
  \bibinfo{person}{Haotian Tang}, \bibinfo{person}{Qinghao Hu},
  \bibinfo{person}{Guangxuan Xiao}, \bibinfo{person}{Jiaming Tang},
  \bibinfo{person}{Yujun Lin}, \bibinfo{person}{Zhijian Liu},
  \bibinfo{person}{Yao Lu}, {and} \bibinfo{person}{Song Han}.}
  \bibinfo{year}{2025}\natexlab{}.
\newblock \bibinfo{title}{LServe: Efficient Long-sequence LLM Serving with
  Unified Sparse Attention}.
\newblock
\showeprint[arxiv]{2502.14866}~[cs.CL]
\urldef\tempurl%
\url{https://arxiv.org/abs/2502.14866}
\showURL{%
\tempurl}


\bibitem[Yang et~al\mbox{.}(2016)]%
        {caption-generation}
\bibfield{author}{\bibinfo{person}{Zhilin Yang}, \bibinfo{person}{Ye Yuan},
  \bibinfo{person}{Yuexin Wu}, \bibinfo{person}{William~W Cohen}, {and}
  \bibinfo{person}{Russ~R Salakhutdinov}.} \bibinfo{year}{2016}\natexlab{}.
\newblock \showarticletitle{Review networks for caption generation}.
\newblock \bibinfo{journal}{\emph{Advances in neural information processing
  systems}}  \bibinfo{volume}{29} (\bibinfo{year}{2016}).
\newblock


\bibitem[Yin et~al\mbox{.}(2024)]%
        {multimodal-survey}
\bibfield{author}{\bibinfo{person}{Shukang Yin}, \bibinfo{person}{Chaoyou Fu},
  \bibinfo{person}{Sirui Zhao}, \bibinfo{person}{Ke Li}, \bibinfo{person}{Xing
  Sun}, \bibinfo{person}{Tong Xu}, {and} \bibinfo{person}{Enhong Chen}.}
  \bibinfo{year}{2024}\natexlab{}.
\newblock \showarticletitle{A Survey on Multimodal Large Language Models}.
\newblock \bibinfo{journal}{\emph{National Science Review}}
  (\bibinfo{date}{Nov.} \bibinfo{year}{2024}).
\newblock
\showISSN{2053-714X}
\urldef\tempurl%
\url{https://doi.org/10.1093/nsr/nwae403}
\showDOI{\tempurl}


\bibitem[Yue et~al\mbox{.}(2023)]%
        {lawllm}
\bibfield{author}{\bibinfo{person}{Shengbin Yue}, \bibinfo{person}{Wei Chen},
  \bibinfo{person}{Siyuan Wang}, \bibinfo{person}{Bingxuan Li},
  \bibinfo{person}{Chenchen Shen}, \bibinfo{person}{Shujun Liu},
  \bibinfo{person}{Yuxuan Zhou}, \bibinfo{person}{Yao Xiao},
  \bibinfo{person}{Song Yun}, \bibinfo{person}{Xuanjing Huang},
  {et~al\mbox{.}}} \bibinfo{year}{2023}\natexlab{}.
\newblock \showarticletitle{Disc-lawllm: Fine-tuning large language models for
  intelligent legal services}.
\newblock \bibinfo{journal}{\emph{arXiv preprint arXiv:2309.11325}}
  (\bibinfo{year}{2023}).
\newblock


\bibitem[Yun et~al\mbox{.}(2021)]%
        {spectr}
\bibfield{author}{\bibinfo{person}{Boxiang Yun}, \bibinfo{person}{Yan Wang},
  \bibinfo{person}{Jieneng Chen}, \bibinfo{person}{Huiyu Wang},
  \bibinfo{person}{Wei Shen}, {and} \bibinfo{person}{Qingli Li}.}
  \bibinfo{year}{2021}\natexlab{}.
\newblock \showarticletitle{Spectr: Spectral transformer for hyperspectral
  pathology image segmentation}.
\newblock \bibinfo{journal}{\emph{arXiv preprint arXiv:2103.03604}}
  (\bibinfo{year}{2021}).
\newblock


\bibitem[Zhang et~al\mbox{.}(2024)]%
        {mmllm}
\bibfield{author}{\bibinfo{person}{Duzhen Zhang}, \bibinfo{person}{Yahan Yu},
  \bibinfo{person}{Jiahua Dong}, \bibinfo{person}{Chenxing Li},
  \bibinfo{person}{Dan Su}, \bibinfo{person}{Chenhui Chu}, {and}
  \bibinfo{person}{Dong Yu}.} \bibinfo{year}{2024}\natexlab{}.
\newblock \bibinfo{title}{MM-LLMs: Recent Advances in MultiModal Large Language
  Models}.
\newblock
\showeprint[arxiv]{2401.13601}~[cs.CL]
\urldef\tempurl%
\url{https://arxiv.org/abs/2401.13601}
\showURL{%
\tempurl}


\bibitem[Zhang et~al\mbox{.}(2023a)]%
        {adalora}
\bibfield{author}{\bibinfo{person}{Qingru Zhang}, \bibinfo{person}{Minshuo
  Chen}, \bibinfo{person}{Alexander Bukharin}, \bibinfo{person}{Nikos
  Karampatziakis}, \bibinfo{person}{Pengcheng He}, \bibinfo{person}{Yu Cheng},
  \bibinfo{person}{Weizhu Chen}, {and} \bibinfo{person}{Tuo Zhao}.}
  \bibinfo{year}{2023}\natexlab{a}.
\newblock \showarticletitle{Adalora: Adaptive budget allocation for
  parameter-efficient fine-tuning}.
\newblock \bibinfo{journal}{\emph{arXiv preprint arXiv:2303.10512}}
  (\bibinfo{year}{2023}).
\newblock


\bibitem[Zhang et~al\mbox{.}(2023b)]%
        {writing-alpaca}
\bibfield{author}{\bibinfo{person}{Yue Zhang}, \bibinfo{person}{Leyang Cui},
  \bibinfo{person}{Deng Cai}, \bibinfo{person}{Xinting Huang},
  \bibinfo{person}{Tao Fang}, {and} \bibinfo{person}{Wei Bi}.}
  \bibinfo{year}{2023}\natexlab{b}.
\newblock \showarticletitle{Multi-task instruction tuning of llama for specific
  scenarios: A preliminary study on writing assistance}.
\newblock \bibinfo{journal}{\emph{arXiv preprint arXiv:2305.13225}}
  (\bibinfo{year}{2023}).
\newblock


\bibitem[Zheng et~al\mbox{.}(2024)]%
        {sglang}
\bibfield{author}{\bibinfo{person}{Lianmin Zheng}, \bibinfo{person}{Liangsheng
  Yin}, \bibinfo{person}{Zhiqiang Xie}, \bibinfo{person}{Chuyue Sun},
  \bibinfo{person}{Jeff Huang}, \bibinfo{person}{Cody~Hao Yu},
  \bibinfo{person}{Shiyi Cao}, \bibinfo{person}{Christos Kozyrakis},
  \bibinfo{person}{Ion Stoica}, \bibinfo{person}{Joseph~E. Gonzalez},
  \bibinfo{person}{Clark Barrett}, {and} \bibinfo{person}{Ying Sheng}.}
  \bibinfo{year}{2024}\natexlab{}.
\newblock \bibinfo{title}{SGLang: Efficient Execution of Structured Language
  Model Programs}.
\newblock
\showeprint[arxiv]{2312.07104}~[cs.AI]
\urldef\tempurl%
\url{https://arxiv.org/abs/2312.07104}
\showURL{%
\tempurl}


\bibitem[Zhou et~al\mbox{.}(2023)]%
        {ifeval}
\bibfield{author}{\bibinfo{person}{Jeffrey Zhou}, \bibinfo{person}{Tianjian
  Lu}, \bibinfo{person}{Swaroop Mishra}, \bibinfo{person}{Siddhartha Brahma},
  \bibinfo{person}{Sujoy Basu}, \bibinfo{person}{Yi Luan},
  \bibinfo{person}{Denny Zhou}, {and} \bibinfo{person}{Le Hou}.}
  \bibinfo{year}{2023}\natexlab{}.
\newblock \bibinfo{title}{Instruction-Following Evaluation for Large Language
  Models}.
\newblock
\showeprint[arxiv]{2311.07911}~[cs.CL]
\urldef\tempurl%
\url{https://arxiv.org/abs/2311.07911}
\showURL{%
\tempurl}


\end{thebibliography}

\newpage

\appendix
\section{Artifact Appendix}

\subsection{Abstract}


This artifact provides a complete workflow to reproduce the performance evaluation of EdgeLoRA, a multi-tenant LLM serving system optimized for edge devices. The source code is publicly available at \url{https://github.com/shenzheyu/EdgeLoRA.git}, with precompiled binaries hosted online for convenience. The artifact supports deployment on Jetson AGX Orin, Jetson Orin Nano, and Raspberry Pi 5 devices, and includes detailed installation and execution instructions.

To reproduce the results in the paper, users can launch the EdgeLoRA server with two arguments specifying the model and the number of LoRA adapters, followed by an experiment script that simulates synthetic workloads. Metrics such as throughput, average request latency, first-token latency, and SLO attainment are automatically reported. The default experiment replicates the results presented in the paper, while the setup can be easily customized via parameters such as request rate, adapter count, and input/output lengths. The full experiment completes within minutes and requires approximately 20GB of disk space.

By following the provided steps, users can replicate the benchmark results or conduct customized experiments to evaluate EdgeL-oRA’s scalability and efficiency across a wide range of configurations.

\subsection{Artifact check-list (meta-information)}


{\small
\begin{itemize}
  \item {\bf Compilation: }  \texttt{gcc/g++}, \texttt{nvcc}
  \item {\bf Binary: } Precompiled EdgeLoRA binary available at \url{https://github.com/shenzheyu/EdgeLoRA/releases/tag/v1.0.0}
  \item {\bf Hardware: } Jetson Agx Orin Developer Kit, Jetson Orin Nano, and Rasperry Pi 5
  \item {\bf Metrics: } Throughput, average request latency, average first-token latency, and SLO attainment.
  \item {\bf Output: } Printed summary of performance metrics to terminal
  \item {\bf Experiments: } Described below
  \item {\bf How much disk space required (approximately)?: } 20GB
  \item {\bf How much time is needed to prepare workflow (approximately)?: } 1 hour
  \item {\bf How much time is needed to complete experiments (approximately)?: } 10 minutes per configuration
  \item {\bf Publicly available?: } Yes
  \item {\bf Code licenses (if publicly available)?: } MIT License
  \item {\bf Archived (provide DOI)?: } 10.6084/m9.figshare.28675676
\end{itemize}
}

\subsection{Description}

\subsubsection{How to access}

\begin{itemize}
    \item Source code: \url{https://github.com/shenzheyu/EdgeLoRA.git}
    \item Binary release: \url{https://github.com/shenzheyu/EdgeLoRA/releases/tag/v1.0.0}
\end{itemize}

\subsubsection{Hardware dependencies}

To match the experiment setup described in the paper, EdgeLoRA is evaluated on the following devices:
\begin{itemize}
    \item Jetson AGX Orin Developer Kit
    \item Jetson Orin Nano
    \item Raspberry Pi 5
\end{itemize}

\subsubsection{Software dependencies}

\begin{itemize}
    \item Ubuntu 22.04
    \item L4T Driver Package Version: 36.6.3
    \item JetPack Version: 6.2
    \item \texttt{g++} Compiler: 11.4.0
    \item Node.js: 20.18.3
\end{itemize}

\subsection{Installation}


The following steps describe how to install EdgeLoRA from source:

\begin{lstlisting}[style=bashstyle, language=bash]
# clone the EdgeLoRA repository
git clone https://github.com/shenzheyu/EdgeLoRA.git

# compile the source code
cd EdgeLoRA/edgelora
export GGML_CUDA=1 # enable CUDA if device has a GPU
make llama-server

# download pre-trained models and adapters
pip install gdown
gdown https://drive.google.com/uc\?id\=1cyU2MUe8V4bo4IuKZG7cEZ2wpxJzD0nn
tar -xzvf models.tar.gz

# install the dependencies of experiment script
cd llama-client
npm install gamma progress

\end{lstlisting}

\subsection{Experiment workflow}

To reproduce the default experiment for EdgeLoRA using the Llama3.1-8B model and 20 LoRA adapters:

\begin{lstlisting}[style=bashstyle, language=bash]
# launch the EdgeLoRA server
bash server.sh Llama3.1-8B 20

# run the default experiment script
cd llama-client
node edge_lora.js

\end{lstlisting}

The script prints the resulting throughput, average request latency, first-token latency, and SLO attainment directly to the terminal.

\subsection{Evaluation and expected results}


The server should launch successfully, and the initial terminal output should indicate that all slots are idle. After running the experiment, performance metrics should be displayed. These results are expected to be consistent with the values reported in the paper, validating the correctness of the artifact setup.

\subsection{Experiment customization}

The server can be launched using the following command with two arguments: \texttt{bash server.sh <model> <lora\_count>}.

\begin{itemize}
    \item \textbf{model}: Specifies the name of the base language model to be served. Supported options include \texttt{OpenELM-1.1B}, \texttt{Llama3.2-3B}, and \texttt{Llama3.1-8B}.
    \item \textbf{lora\_count}: Indicates the total number of LoRA adapters to be managed by the server. This value can range from a few dozen to several thousand.
\end{itemize}

The above experiment script could also be customized with multiple arguments in the `llama-client/edge\_lora.js` file:

\begin{itemize}
    \item \textbf{n}: Number of LoRA adapters available in the system. Controls adapter diversity.
    \item \textbf{alpha}: Power-law exponent that defines the skewness of request distribution across adapters.
    \item \textbf{R}: Total request rate, i.e., how many requests per second are sent across all adapters.
    \item \textbf{cv}: Coefficient of variance for arrival intervals in the Gamma process, defining burstiness of the workload.
    \item \textbf{traceDuration}: Duration of the synthetic trace (in milliseconds), default representing 5 minutes.
    \item \textbf{Il, Iu}: Lower and upper bounds for input token lengths sampled from a uniform distribution.
    \item \textbf{Ol, Ou}: Lower and upper bounds for output token lengths, also sampled uniformly.
\end{itemize}

\subsection{Notes}

\begin{itemize}
    \item The \texttt{edgelora\_wo\_aas} folder contains the implementation of EdgeLoRA without adaptive adapter selection. Its usage is similar to the standard EdgeLoRA workflow.
    
    \item The \texttt{adapter-router} folder provides the implementation for fine-tuning and evaluating the adapter router. This component requires a custom version of the HuggingFace Transformers library, which can be installed using:

    \begin{lstlisting}[style=bashstyle, language=bash]
pip install git+https://github.com/shenzheyu/transformers.git@edgelora#egg=transformers
    \end{lstlisting}
\end{itemize}

\end{document}